\documentclass[twocolumn]{aastex62}
\usepackage[utf8]{inputenc}
\usepackage{natbib}
\usepackage{amsmath, mathtools}
\usepackage{hyperref}
\usepackage{color}
\usepackage{comment}
\usepackage[T1]{fontenc}
\usepackage{gensymb}
\usepackage[caption=false]{subfig}

\usepackage{subfig}
\usepackage{graphics}
\usepackage[mathlines]{lineno}

\graphicspath{{./}}

\shorttitle{WR Binaries at the Galactic center}
\shortauthors{Bentley et al.}

\begin{document}

\title{A Kinematic Study of Wolf-Rayet Stars at the Galactic Center I: Binary Candidates and Constraints on the Binary Fraction}

\author[0000-0001-7017-8582]{Rory O. Bentley}
\affil{Department of Physics and Astronomy UCLA, Los Angeles, CA 90095-1547, USA}

\author[0000-0001-9554-6062]{Tuan Do}
\affil{Department of Physics and Astronomy UCLA, Los Angeles, CA 90095-1547, USA}

\author[0000-0003-3230-5055]{Andrea Ghez}
\affil{Department of Physics and Astronomy UCLA, Los Angeles, CA 90095-1547, USA}

\author[0000-0003-3765-8001]{Devin Chu}
\affil{Department of Physics and Astronomy UCLA, Los Angeles, CA 90095-1547, USA}
\affil{'Imiloa Astronomy Center of Hawai'i,
University of Hawai'i Hilo, Hilo, HI 96720, USA}

\author[0000-0001-5800-3093]{Anna Ciurlo}
\affil{Department of Physics and Astronomy UCLA, Los Angeles, CA 90095-1547, USA}

\author[0000-0002-2836-117X]{Abhimat K. Gautam}
\affil{Department of Physics and Astronomy UCLA, Los Angeles, CA 90095-1547, USA}

\author[0009-0004-0026-7757]{Zoë Haggard}
\affil{Department of Physics and Astronomy UCLA, Los Angeles, CA 90095-1547, USA}

\author[0000-0003-2874-1196]{Matthew W. Hosek Jr.}
\altaffiliation{Brinson Prize Fellow}
\affil{Department of Physics and Astronomy UCLA, Los Angeles, CA 90095-1547, USA}

\author[0000-0003-2400-7322]{Kelly Kosmo O'neil}
\affil{Department of Physics and Astronomy UCLA, Los Angeles, CA 90095-1547, USA}
\affil{Department of Physics, 
University of Nevada Reno
Reno, NV 89557, USA}

\author[0000-0001-7003-0588]{Rebecca Lewis-Merrill}
\affil{Department of Physics and Astronomy UCLA, Los Angeles, CA 90095-1547, USA}

\author{Gregory D. Martinez}
\affil{Department of Physics and Astronomy UCLA, Los Angeles, CA 90095-1547, USA}

\author[0000-0002-3081-8597]{Anna Pusack}
\affil{Department of Astronomy,
University of California Berkeley, Berkeley, CA, USA}

\author[0000-0001-5972-663X]{Shoko Sakai}
\affil{Department of Physics and Astronomy UCLA, Los Angeles, CA 90095-1547, USA}

\author[0000-0001-9611-0009]{Jessica R. Lu}
\affil{Department of Astronomy,
University of California Berkeley, Berkeley, CA, USA}

\author[0000-0002-6753-2066]{Mark R. Morris}
\affil{Department of Physics and Astronomy UCLA, Los Angeles, CA 90095-1547, USA}


\author{Keith Matthews}
\affil{Division of Physics, Mathematics and Astronomy, Pasadena, CA 91125, USA}

\correspondingauthor{Rory Bentley}
\email{rbentley@astro.ucla.edu}

\begin{abstract}
We report the binary fraction of Wolf-Rayet (WR) stars within 0.5~pc of the Galactic center obtained through the longest time-baseline (1994-2024) kinematic study of this population of stars. The new radial velocity ($v_{z}$) data we present is primarily from the W. M. Keck Observatory, with additional $v_{z}$ measurements from Gemini North Observatory. When combining our new $v_{z}$ measurements with literature measurements, we find $v_{z}$ variations suggesting the presence of a companion for five out of 27 WR stars, of which two are newly identified here (IRS~13E4, S8-181), along with three previously detected binaries (IRS~16SW, IRS~16NE, S4-258). Based on our experimental sensitivity and expected properties of the underlying population, we infer the binary fraction of the WR stars in the Galactic center to be 0.56$\pm$0.18. This is consistent with previous photometric studies of the young stars in the Galactic center, and with the binary fraction of field WR stars. When our results are combined with the results of previous photometric work, we find a binary fraction of 0.69$\pm$0.17 for the WR stars in the Galactic center.

\end{abstract}


\section{Introduction}

\begin{figure*}
\begin{center}
\includegraphics[width=7in]{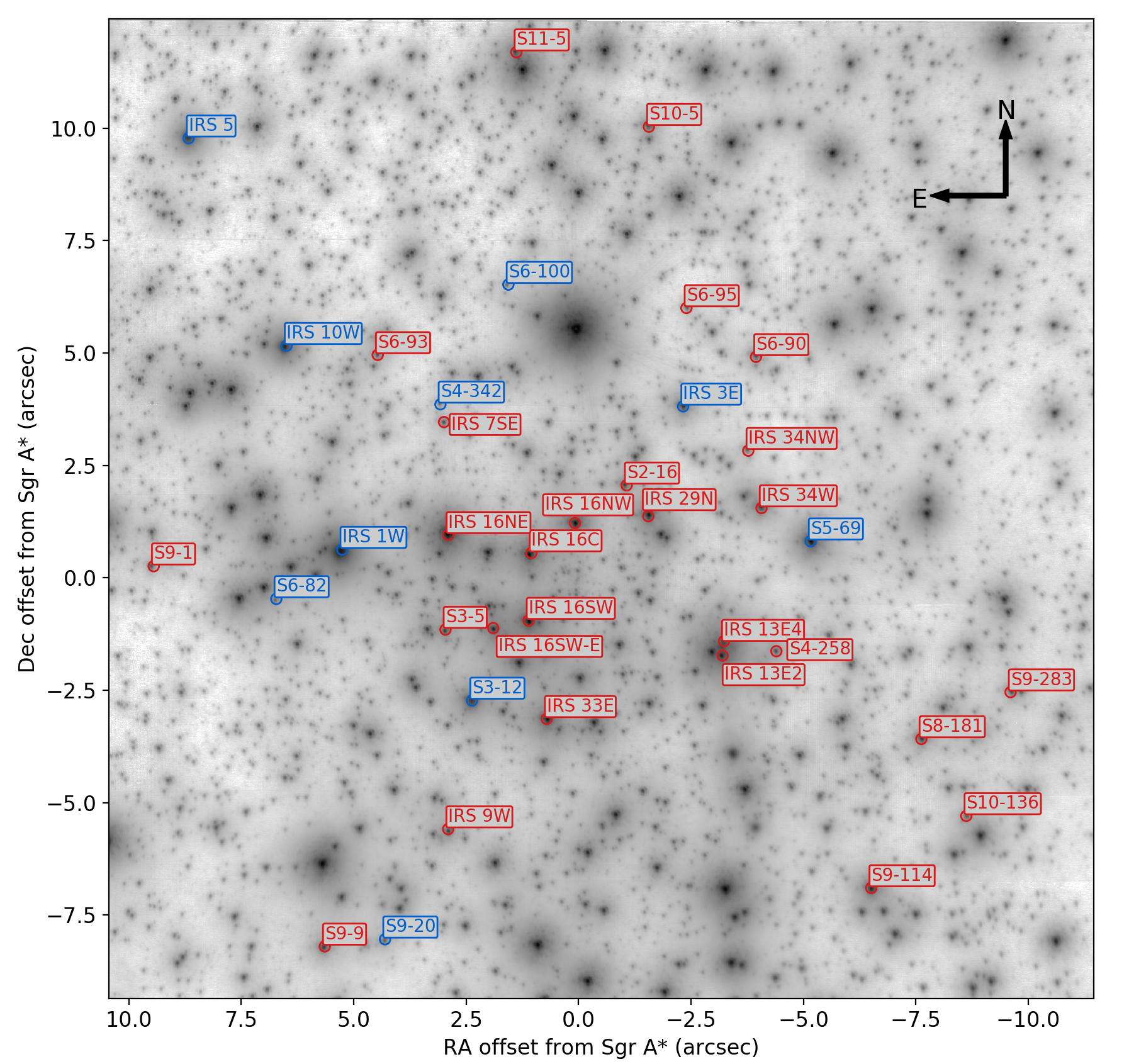}
\caption{\footnotesize Location of all the known Wolf-Rayet (WR) stars and WR star candidates in the Galactic center's nuclear star cluster. Stars encircled in red are members of the binary search sample and those encircled in blue are the remaining WR stars or WR star candidates. Additional names, spectral types, and other information for the WR stars are provided in Table~\ref{tab:wr_table}. The points are overlaid on a K-band adaptive-optics image obtained on 2006-05-03 (UT) with the NIRC2 instrument at the W. M. Keck Observatory.}3
\label{fig:wr_map}
\end{center}
\end{figure*}

The presence of Wolf-Rayet (WR) stars at the Galactic center presents the unique opportunity to study how evolved massive stars interact with a supermassive black hole (SMBH) and its environment. Around 30 WR stars have been identified within $\sim$0.5~pc of Sgr~A* (Figure~\ref{fig:wr_map}), indicating that the Galactic center hosts one of the highest densities of these stars known \citep{VanDerHucht2006}. Owing to their high luminosities ($>10^{5}~L_{\odot}$, \citet{Martins2007}), WR stars were some of the first stars to be identified in the Galactic center in early spectroscopic studies \citep{Forrest1987,Allen1990,Krabbe1991, Najarro1994, Najarro1997}. The advent of adaptive-optics-assisted integral-field spectroscopy helped to complete the census of WR stars \citep{Paumard2006, Bartko2009, Tanner2005}. Medium-resolution (R $\sim$ 4000) K and H-band spectra of the atmospheres of these stars showed that they range in mass from 10 to 82~$M_{\odot}$, with winds of 450 -- 2500~km/s, and mass loss rates of 10$^{-5.3}$--10$^{-3.95}$~$M_{\odot}$/yr \citep{Martins2007}.

The origin of the young stellar population in the Galactic center, including the WR stars, remains poorly understood. It was expected that the tidal shear from Sgr~A* would prevent any stars from forming at their current locations, and the densities required to overcome the tidal shear are many orders of magnitude higher than the observed density at the present time \citep{Morris1993, Ghez2003,Alexander2005,Jackson1993,Christopher2005,MonteroCastano2009}. However, there have been star-formation models proposed in which the tidal forces can presumably be overcome, such as star formation in an accretion disk around Sgr~A*, which can also explain the observed clockwise disk of young stars \citep{Levin2003, Nayakshin2007, Bonnell2008}. Simulations of collapsing accretion disks can produce gas at the densities required to overcome the tidal shear from Sgr~A* at the distances from the SMBH at which the stellar disk is observed \citep[0.1 -- 0.5~pc,][]{Levin2003,Nayakshin2007,Bonnell2008}. 

One angle from which to probe star formation is through measurement of the binary fraction, as it can provide a strong test of star formation models \citep[e.g.][]{Duchene2013}. For example, star formation models in an accretion disk predict a higher binary fraction than typical star formation, as a result of the relatively high fragmentation rate in a disk of gaseous structures resulting from compressional shocks and/or local self-gravity \citep{Nayakshin2007}. 

The binary fraction can also probe the dynamical evolution of the young stars at the Galactic center after they have formed, through mergers and evaporation of binaries through three-body interactions with Sgr A* or with surrounding stars \citep[e.g.][]{Naoz2016,Stephan2016,Rose2020}. The presence of binaries can also affect the velocity measurements that are used to determine the membership of stars in the different kinematic structures at the Galactic center, which can in turn lead to underestimates of the number of stars in these structures \citep{Naoz2018}. 

While the WR star population in the Galactic center's central parsec is well-known, its binary fraction is not. Prior to the study reported in this paper, there were three known WR star binaries known in this region, but the studies which found them either focused on a specific star or focused on a smaller sample of WR stars (8 stars or fewer) than the 27 targeted in this work \citep{Pfuhl2014, Martins2006, Moultaka2005}. This limited their ability to constrain population-level characteristics. 

The binary fraction of the whole young stellar population in the Galactic center is estimated to be greater than $71\%$ based on photometric studies of eclipsing binaries \citep{Gautam2024}. However, spectroscopic studies are more sensitive to the presence of binaries, since they allow detection of non-eclipsing systems otherwise missed in photometric experiments. The most complete spectroscopic study so far led to an inferred spectroscopic binary fraction of $0.30^{+0.34}_{-0.21}$ \citep{Pfuhl2014}. However, how this translates into an intrinsic binary fraction is unclear. They also targeted 12 particularly bright young stars rather than specific spectral types such as the WR stars (8 WR stars were included in the \citealt{Pfuhl2014} sample). \citet{Chu2023} searched for spectroscopic binaries among 16 young main-sequence B stars within 0.04~pc of Sgr A* and found none, while \citet{Gautam2024} found a binary fraction consistent with field populations in a photometric search targeting 102 young stars (including both WR and OB stars) out to $\sim$0.4~pc from Sgr A*. These results together suggest a radial increase in the binary fraction away from Sgr A*.  

WR stars in particular are well-suited for spectroscopic binary searches. Their high luminosities ($10^{5}-10^{6} L_{\odot}$) make them very easy to observe compared to the rest of the Galactic center young stars, and as high-mass stars have high binary fractions \citep[e.g.,][]{Sana2012,Dsilva2020,Dsilva2022,Dsilva2023,Offner2023,Shenar2024}.

In this work, we assemble a sample of 27 Galactic center WR stars (Section \ref{sect:wr_sample}), using both new and literature radial velocity measurements (Section \ref{sect:rv_measurement}). We then conduct a search for radial velocity variations and locate two new binary candidates among the WR stars (S8-181, IRS~13E4, Section \ref{sect:rv_measurement}), allowing us to estimate the binary fraction of the Galactic center WR stars (Section~\ref{sec:discussion}).

\section{Wolf-Rayet star sample}    
\label{sect:wr_sample}
We assembled the most complete sample of WR stars and WR star candidates within 0.5~pc of Sgr~A*. Assembling the sample began with the WR star catalog from \citet{VanDerHucht2006}, which contains 43 stars classified as WR stars within 0.5 pc of Sgr A*. Six objects in the original catalog are not consistent with being WR stars \citep{Krabbe1995, Paumard2006, Sanchez-Bermudez2014, Tanner2005, Tanner2002, Feldmeier-Krause2015, Moultaka2005,Clenet2001, Moultaka2009,Fritz2010,Zhu2020}. This produces a catalog of 37 WR stars or candidates; Table~\ref{tab:wr_table} provides their names, locations, and spectral types.

The WR stars fall into two broad spectral types: WN and WC. WN stars have spectra with strong N and He lines, and are thought to be the precursors to WC stars, which have spectra dominated by C and He lines. These two spectral types have different wind properties, with WN having more variation in time, with wind-driven variations of up to 50 km/s for WN stars versus 10 km/s for WC stars \citep{Dsilva2020,Dsilva2022,Dsilva2023}. We include Ofpe/WN9 stars as WN stars, which are thought to be stars transitioning from the Luminous Blue Variable (LBV) evolutionary state to the more evolved WR stage \citep{Martins2007}. In total, there are 18 WN and 9 WC stars in the full catalog.

For the binary study, we only use stars with at least two radial velocity measurements.  Our repeated measurements cover 14 stars and the remaining 13 stars have multiple radial velocities reported in the literature (see Section \ref{sect:rv_measurement}) for a total of 27 stars for this investigation. Many of the stars that are not included in the binary search sample are WC stars surrounded by dust, with featureless spectra \citep{Tanner2005,Feldmeier-Krause2015,Sanchez-Bermudez2014}. Our WR star binary search sample triples the number of WR stars included in previous spectroscopic binary WR star searches in the Galactic center \citep{Tanner2006, Martins2006, Pfuhl2014}.

Our final sample includes 18 WN stars, and 9 WC stars. Many of the stars which are not included in the binary search sample are WC stars surrounded by dust, with featureless spectra \citep{Tanner2005,Feldmeier-Krause2015,Sanchez-Bermudez2014}. Table~\ref{tab:wr_table} provides the names, locations, and spectral types from the literature of the WR stars.

\begin{deluxetable*}{lllcccllcccc}
\tablecolumns{12} 
\tablewidth{0pc}
\tabletypesize{\scriptsize}
\tablecaption{Sample of Wolf-Rayet stars in the Nuclear Star Cluster\label{tab:wr_table}}
\tablehead{ 
	\colhead{Name} &
	\colhead{Aliases$^{a}$} &
	\colhead{Kp} &
	\colhead{$\Delta$RA$^{b}$} &
	\colhead{$\Delta$Dec$^{b}$} &
    \colhead{$R_{2D}$} &
	\colhead{$T_{0}$} &
	\colhead{Sp. Type} &
	\colhead{N of new $v_{z}$} &
	\colhead{N of new $v_{z}$ used} &
	\colhead{N of lit. $v_{z}$ used} &
	\colhead{In search?} \\
    \colhead{} &
    \colhead{} &
	\colhead{} &
    \colhead{arcsec} &
    \colhead{arcsec} &
    \colhead{arcsec} &
	\colhead{} &
 	\colhead{} &
	\colhead{Kn3/K} &
 	\colhead{} &
 	\colhead{} &
 	\colhead{} 
}
\startdata
IRS 16C & S96$^{1}$,E20$^{2}$ &  9.9 & 1.05 & 0.55 & 1.19 & 2009.99 & Ofpe/WN9 & 47/11 & 58 & -- & Yes\\ 
IRS 16NW & E19$^{2}$ & 10.1 & 0.08 & 1.22 & 1.22 & 2010.05 & Ofpe/WN9 & 22/9 & 31 & -- & Yes \\ 
IRS 16SW$^{d}$ & S97$^{1}$,IRS 16SW(W)$^{3}$,E23$^{2}$ & 10.0 & 1.11 & -0.95 & 1.46 & 2009.82 & Ofpe/WN9 & 0/10 & 10 & -- & Yes\\ 
IRS 29N & E31$^{2}$  & 10.4 & -1.55 & 1.37 & 2.07 & 2010.16 & WC9 & 0/0 & -- & 2 & Yes\\ 
IRS 16SW-E & E32$^{2}$,MPE +1.6-6.8$^{4}$ & 11.1 & 1.89 & -1.12 & 2.19 & 2010.05 & WC8/9 & 0/0 & -- & 2 & Yes\\
S2-16 & E35$^{2}$,MPE-1.0-3.5$^{4}$ & 12.0 & -1.07 & 2.06 & 2.29 & 2010.2 & WC8/9 & 0/0 & -- & 2  & Yes\\ 
IRS 16NE$^{d}$ & E39$^{2}$ & 9.1 & 2.89 & 0.95 & 3.04 & 2010.03 & Ofpe/WN9 & 3/0 & -- & 5 & Yes \\ 
S3-5 & E40$^{2}$,IRS 16SE2$^{5}$ & 11.8 & 2.96 & -1.15 & 3.18 & 2009.91 & WN5/6 & 5/1 & 6 & -- & Yes \\ 
IRS 33E & E41$^{2}$ & 10.1 & 0.71 & -3.14 & 3.22 & 2010.18 & Ofpe/WN9 & 3/1 & -- & 7 & Yes \\ 
IRS 13E4 & E48$^{2}$ & 11.7 & -3.23 & -1.41 & 3.50 & 2009.36 & WC9 & 5/1 & 6 & -- & Yes \\ 
IRS 13E2 & E51$^{2}$ & 10.6 & -3.20 & -1.73 & 3.64 & 2009.98 & WN8 & 5/3 & 7 & -- & Yes \\ 
IRS 34W & E56$^{2}$ & 11.6 & -4.07 & 1.55 & 4.36 & 2010.19 & Ofpe/WN9 & 5/0 & 5 & -- & Yes \\ 
IRS 7SE & E59$^{2}$, [PMM2001] B9$^{6}$ & 13.4 & 2.99 & 3.46 & 4.57 & 2009.59 & WC9 & 5/0 & 3 & -- & Yes \\ 
S4-258$^{d}$ & E60$^{2}$ & 12.6 & -4.40 & -1.63 & 4.69 & 2009.76 & WN7? & 0/0 & -- & 16$^{e}$ & Yes \\ 
IRS 34NW & E61$^{2}$ & 13.2 & -3.78 & 2.83 & 4.72 & 2010.28 & WN7 & 3/0 & 3 & -- & Yes \\ 
IRS 9W & E65$^{2}$ & 12.1 & 2.89 & -5.59 & 6.29 & 2010.25 & WN8 & 0/0 & -- & 5 & Yes \\ 
S6-90 & E66$^{2}$,IRS 7SW$^{5}$,WR1$^{8}$ & 12.5 & -3.95 & 4.91 & 6.30 & 2010.14 & WN8/WC9 & 0/0 & -- & 2 & Yes \\ 
S6-95 & E68$^{2}$,IRS 7W$^{5}$,WR2$^{8}$ & 13.4 & -2.40 & 6.00 & 6.46 & 2010.18 & WC9 & 0/0 & -- & 7 & Yes \\ 
S6-93 & E70$^{2}$,IRS 7E2$^{5}$,IRS 7ESE$^{5}$ & 12.7 & 4.47 & 4.96 & 6.68 & 2010.18 & WN8 & 0/0 & -- & 4 & Yes \\ 
S8-181 & E74$^{2}$,AFNW$^{7}$ & 11.6 & -7.62 & -3.58 & 8.42 & 2008.18 & WN8 & 2/3 & 5 & -- & Yes \\ 
S9-1 & E78$^{2}$,[PMM2001] B1$^{6}$ & 12.6 & 9.44 & 0.26 & 9.44 & 2010.23 & WC9 & 2/0 & 2 & -- & Yes \\ 
S9-114 & E79$^{2}$,AF$^{7}$ & 11.2 & -6.51 & -6.90 & 9.49 & 2007.76 & Ofpe/WN9 & 1/1 & -- & 3 & Yes \\ 
S9-283 & E81$^{2}$,AFNWNW$^{7}$ & 12.5 & -9.61 & -2.54 & 9.94 & 2007.37 & WN7 & 0/0 & -- & 2 & Yes \\ 
S9-9 & E80$^{2}$,IRS 9SE$^{5}$ & 11.8 & 5.64 & -8.20 & 9.95 & 2009.44 & WC9 & 0/0 & -- & 5 & Yes \\ 
S10-136 & E82$^{2}$,Blum$^{7}$ & 13.0 & -8.62 & -5.29 & 10.11 & 2007.38 & WC8/9 & 0/0 & -- & 2 & Yes \\ 
S10-5 & E83$^{2}$,WR3$^{8}$ & 11.9 & -1.56 & 10.03 & 10.15 & 2010.0 & WN8/WC9 & 2/0 & -- & 6 & Yes \\ 
S11-5 & E88$^{2}$ & 11.9 & 1.38 & 11.69 & 11.77 & 2010.11 & WN8/9 & 5/1 & 6 & -- & Yes \\ 
\hline
S3-12 & IRS 21$^{5}$ & 12.0 & 2.37 & -2.73 & 3.62 & 2003.37 & WC9? & 0/0 & -- & -- & No \\ 
IRS 3E & E58$^{2}$ & 15.0 & -2.33 & 3.81 & 4.46 & 2010.11 & WC5/6 & 0/0 & -- & 1 & No \\
S4-342 & IRS 7SE2$^{5}$ & 14.0 & 3.07 & 3.86 & 4.93 & 2010.65 & WC? & 0/0 & -- & 1 & No \\
S5-69 & IRS 6E$^{5}$ & 10.2 & -5.16 & 0.81 & 5.22 & 2010.5 & WC9? & 0/0 & -- & -- & No \\ 
IRS 1W & E63$^{2}$ & 10.8 & 5.25 & 0.63 & 5.29 & 2011.17 & WCLd? & 0/0 & -- & 1 & No \\ 
S6-100 & E71$^{2}$ & 13.9 & 1.56 & 6.52 & 6.70 & 2010.38 & WC8/9? & 0/0 & -- & 1 & No \\ 
S6-82 & E72$^{2}$ & 13.5 & 6.72 & -0.47 & 6.74 & 2010.2 & WC9? & 0/0 & -- & 1 & No \\ 
IRS 10W & -- & 11.2 & 6.49 & 5.16 & 8.29 & 2009.56 & WCLd? & 0/0 & -- & 1 & No \\ 
S9-20 & E76$^{2}$,IRS 9SW$^{5}$ & 13.1 & 4.30 & -8.04 & 9.12 & 2010.5 & WC9 & 0/0 & -- & 1 & No \\ 
IRS 5 & -- & --$^{c}$ & 8.62 & 9.83 & 13.07 & 2010.426 & WC9? & 0/0 & -- & -- & No \\ 
\enddata
\tablenotetext{a}{Sources: 1) \citet{Eckart1997}, 2) \citet{Paumard2006}, 3) \citet{Zhu2008}, 4) \citet{Eckart1995}, 5) \citet{Bailey1984}, 6) \citet{Paumard2001}, 7) \citet{Genzel1996}, 8) \citet{Moultaka2005}}
\tablenotetext{b}{Offsets from Sgr A*}
\tablenotetext{c}{IRS 5 has a $K_{S}$=11.25 from \citet{Schodel2010}}
\tablenotetext{d}{Previously known binary star from \citet{Ott1999,Tanner2006,Martins2006,Peeples2007,Zhu2008,Pfuhl2014}}
\tablenotetext{e}{We use the number of $v_{z}$ measurements from \citet{Pfuhl2014}, but not the values, which were not reported}

\end{deluxetable*}

\section{Observations and Data Reduction}    
\label{sect:obs_data}

A total of 110 new integral field unit observations (Table \ref{tab:obs_table}) were taken on the Keck and Gemini Observatories, targeting 15 out of the 27 WR stars in the sample.  The majority of these (101) were done with the Keck integral-field spectrograph OSIRIS \citep{Larkin2006} from 2006-2024, using the Kn3 (2.121-2.229~$\mu$m) or Kbb (1.96-2.38$~\mu$m) filters for the OSIRIS observations at a spectral resolution of R = 4000 \citep{Larkin2006}. The remaining observations are from the Gemini North telescope observed in 2018 with NIFS using the K-band (1.965-2.430 $\mu$m) filter at a resolution of R = 5000 \citep{McGregor2003} from program GN-2018A-Q-123 (PI: Do). Table \ref{tab:obs_table} (located in Appendix \ref{sect:obs_and_rvs}) summarizes the details of these observations.

These data were reduced using the OSIRIS Data Reduction Pipeline \citep{Lyke2017,Lockhart2019} and the Nifty4Gemini IFU Pipeline\footnote[1]{\url{https://github.com/mrlb05/nifty4gemini}}.  This processing includes rectification and wavelength calibration to produce data cubes. The uncertainties from the wavelength calibration for both instruments are small (<7 km/s,   \citet{Lockhart2019,Lemoine-Busserolle2019}) compared to our typical $v_{z}$ uncertainties, but nonetheless we add in quadratures 7 km/s for the OSIRIS observations and 6 km/s for the NIFS observations.

\begin{figure*}[tb]
\begin{center}
\includegraphics[width=7in]{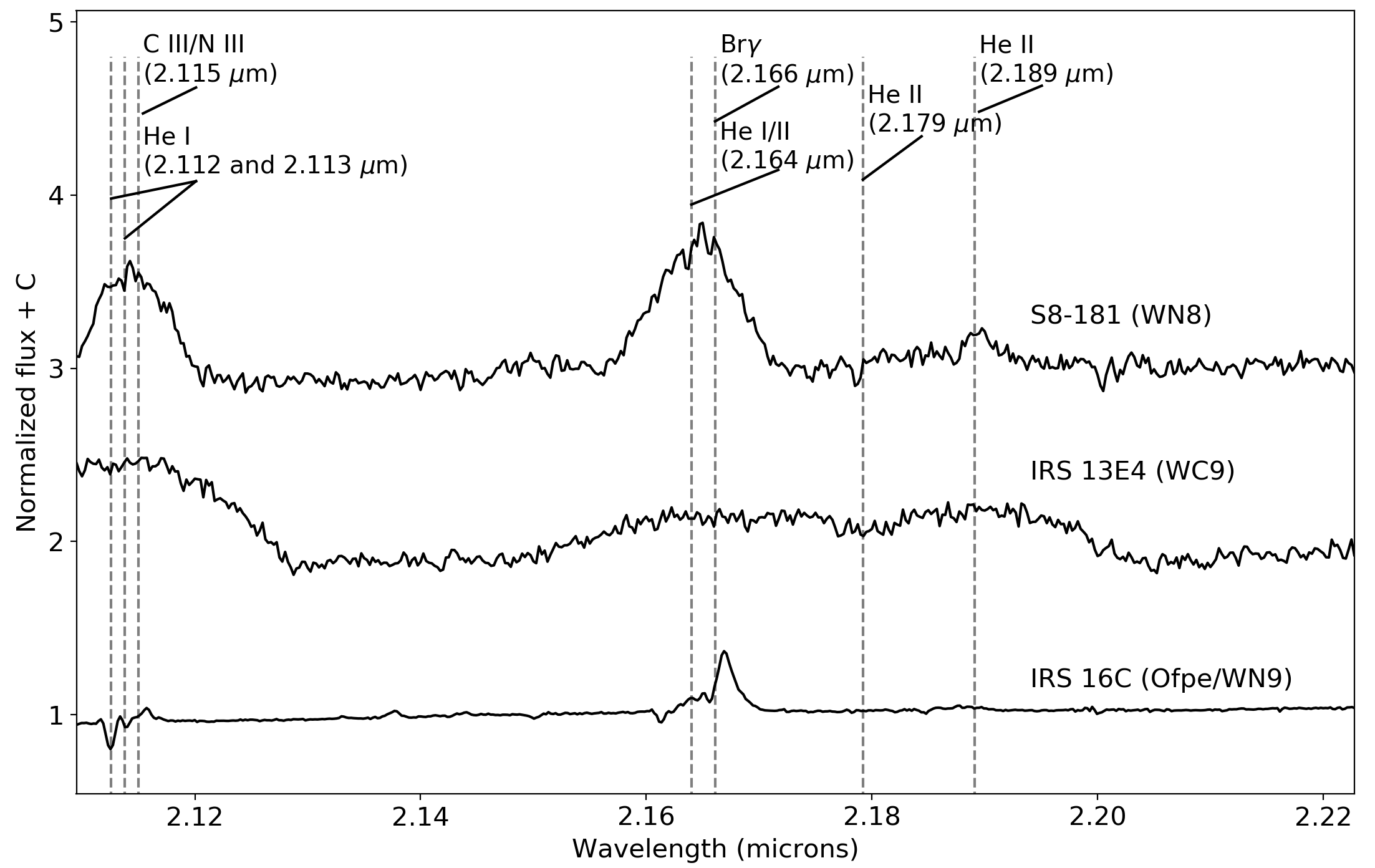}
\caption{\footnotesize Representative spectra for the various WR spectral types in this work, with prominent spectral lines labeled. The spectral type of a WR star determines whether we are able to obtain absolute $v_{z}$ measurements from the 2.112 $\mu$m He I absorption line, or whether we are only able to obtain relative radial velocities from cross-correlation with a template spectrum (e.g. IRS~13E4 and S8-181). The two lines of the He I doublet, which includes the 2.112 $\mu$m line, are marked with dashed lines. Additionally, differing scales of wind-driven $v_{z}$ variability between WN and WC stars \citep{Dsilva2020, Dsilva2022, Dsilva2023} determine the scale of the systematic error applied to account for changing wind structure (WC stars showing less variability vs WN stars). Spectral types for the WR stars are provided in Table~\ref{tab:wr_table}.}
\label{fig:various_wrs}
\end{center}
\end{figure*}

The stellar spectra are extracted using the method described in \citet{Do2009, Do2013}, which results in calibrated one-dimensional spectra. We additionally only include spectra with signal-to-noise ratios greater than 20 to ensure robust radial velocity measurements (see \citet{Do2019}).  This produces 151 spectra over our sample of 27 stars. Example spectra for different WR spectral types are shown in Figure \ref{fig:various_wrs}. For these observations, the median signal-to-noise ratio is 80, and for the brighter stars like IRS~16C it reaches over 200 (Table~\ref{tab:rvs_table_all}).

\section{Radial Velocity Measurements}    
\label{sect:rv_measurement}

\subsection{New $v_{z}$ Measurements} 

\begin{figure}
\includegraphics[width=3.5in]{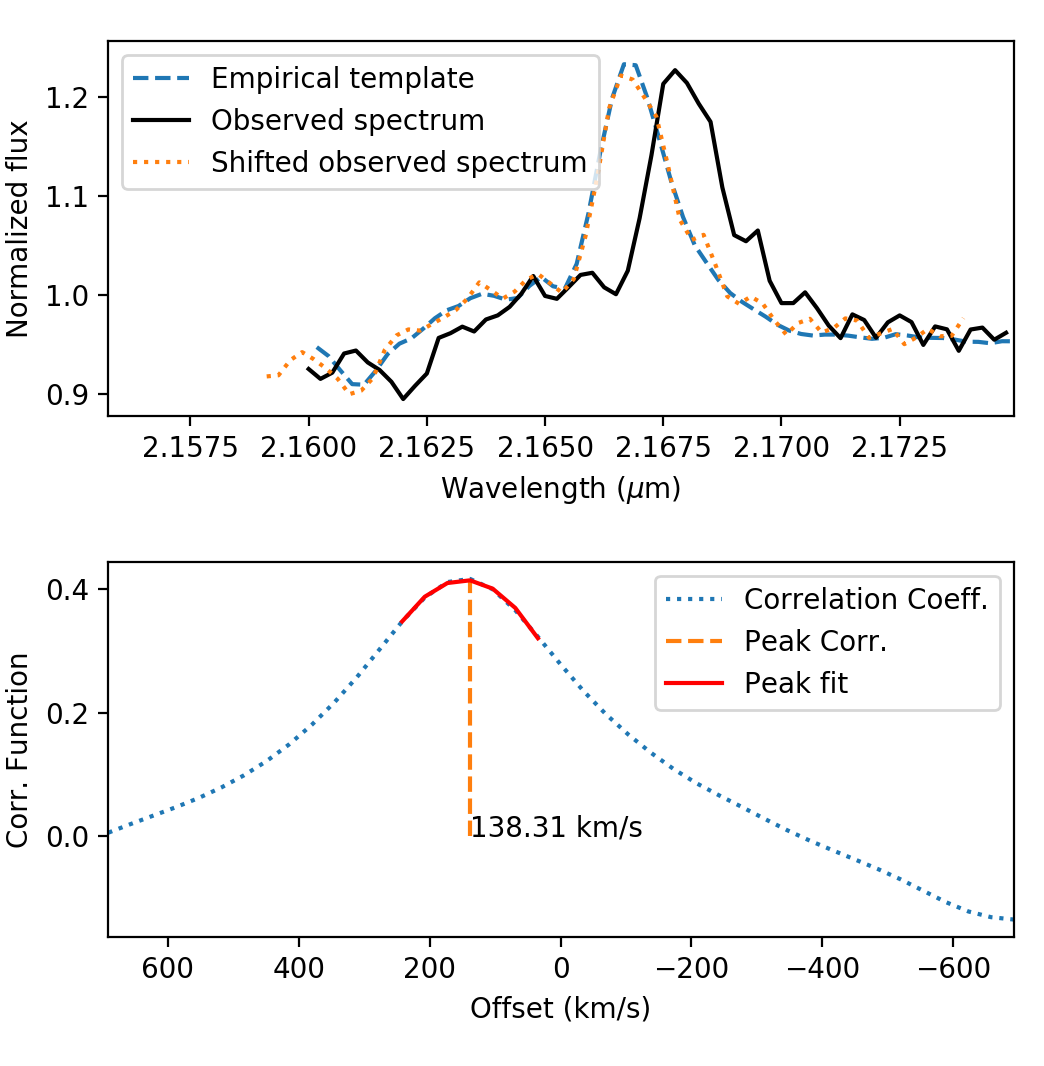}
\caption{\footnotesize A figure showing the cross-correlation method of determining relative $v_z$ values. \textit{Top:} Plot showing an observed spectrum of a star (IRS~16C on 2006-06-18 UT, solid black line), the constructed template for the star (blue dashed line), and the observed spectrum shifted to the velocity of the template after the maximum of the correlation function is identified (orange dotted line).  \textit{Bottom:} A plot of the correlation function (blue dotted line) as a function of offset in pixels for the same spectra. The fit to the peak of the correlation function is plotted (red solid line), along with the location of the maximum and its velocity offset (orange dashed line). }
\label{fig:ccv_reference_fig}
\end{figure}

From our spectra, we obtain relative $v_{z}$ measurements using cross-correlation.  The cross-correlation method requires a window function and a reference spectrum \citep[see e.g.][]{Tonry1979}. We use a top-hat window function over the lines in the spectra with the least wind-driven variations identified in \citet{Dsilva2020,Dsilva2022,Dsilva2023}; the window function range is chosen specifically for each star, as the optimal spectral features vary between stars depending on the spectral type and the strength of the lines (see Table \ref{tab:rvs_table_all} for details on the window used). For this work, we constructed a template for each star using a three-step process. In the first step, we use the highest signal-to-noise spectrum as the template and derive relative $v_{z}$ measurements. Then, a template spectrum was constructed by averaging the observed spectra weighted by their signal-to-noise ratio after shifting them to a common rest wavelength scale using the relative velocities with the observed spectrum template. The cross-correlation was then redone using the newly constructed template spectrum as a reference. See Figure \ref{fig:ccv_reference_fig} for an example of this final step. The change in local standard of rest velocity is then removed from the measured relative velocity, correcting for the Earth's rotation, its motion around the Sun, and the Sun's peculiar motion with respect the the local standard of rest \citep[see also][]{Ghez2008,DehnenBinney1998}. 

This method of binary search and template construction is commonly used in works studying WR binaries \citep[e.g.][]{Dsilva2020,Dsilva2022,Dsilva2023,Shenar2017,Shenar2019,Shenar2021}. Our relative $v_{z}$ measurements are listed in Table~\ref{tab:rvs_table_all}. See Section~\ref{sect:wind_effects} for a discussion of the uncertainties. An advantage of using empirical templates constructed from observed spectra over synthetic templates is that they are not sensitive to the specific choice of parameters that determine line shapes and strengths (e.g., metallicity of the winds).

Cross-correlation to measure $v_{z}$ is a reliable method for binary searches when the spectral features have fairly invariant shapes, as for 14 out of 15 of our sample of WR stars with new spectra. However, IRS 16SW is a double-lined spectroscopic binary, which biases the $v_{z}$ determinations using the cross-correlation method for this star. We therefore determine the $v_{z}$ by fitting the 2.112 $\mu$m He I absorption line using a Gaussian model. In the Ofpe/WN9 stars, this line is generated near the photosphere and so is more indicative of the true $v_{z}$. This method is less precise than cross correlation, but is more accurate for IRS~16SW. \citet{Martins2006}, \citet{Tanner2006}, and \citet{Zhu2008} obtained $v_{z}$ measurements for this star using similar methods to fit the 2.112 $\mu$m He I profile.

\subsection{New $v_{z}$ Uncertainties}    
\label{sect:wind_effects}
The statistical error in the $v_{z}$ measurements was calculated by running Monte Carlo simulations in which random noise with a scatter of one over the signal-to-noise ratio was added to the data before the cross-correlation or Gaussian fitting. 1000 simulations were run for each spectrum, and the scatter in the resulting $v_{z}$ measurements was used as the uncertainty for the relative measurements. Typically, the statistical errors are 13 km/s, but range from 0.8 to 120 km/s. 

We also include a systematic error to account for spectroscopic variability arising from changes in wind structure in our $v_{z}$ measurements. WR stars, in particular late-type WN stars, have been found to have line profile variability \citep{Dsilva2023,St-Louis2009,Chene2011,Chene2020,Michaux2014}, on timescales of days. \citet{Dsilva2020, Dsilva2022, Dsilva2023} estimated the variability of measured $v_{z}$ values using different choices of lines to measure $v_{z}$ across different WR spectral types, and found that relative $v_{z}$ measurements from cross-correlation for single stars can vary by typically 15 km/s for early and late WN stars \citep{Dsilva2022, Dsilva2023}. WC stars on the other hand, have much smaller variations at $\sim$3 km/s \citep{Dsilva2020}. To account for these wind-driven variations, we add 15 km/s in quadrature to the statistical uncertainties for the WN stars (including Ofpe/WN9 and hybrid WN/WC stars), and 3 km/s to the statistical uncertainties for the WC stars. For WN stars, the wind variability generally dominates the $v_{z}$ uncertainty.

\subsection{$v_{z}$ Measurements from the Literature}
\label{sect:lit_data}
We include $v_{z}$ measurements from the literature for the identification of candidate binary stars \citep{Eckart1997,Genzel2000, Tanner2005, Paumard2006, Zhu2008, Moultaka2005}.  We choose to use only those measurements with dates reported to better than monthly precision. 
The final binary search sample includes 70 previously-published $v_{z}$ measurements (Table~\ref{tab:rvs_table_all}). We choose to use $v_{z}$ measurements with the date specified to the nearest month to be included in the binary search. Additionally, because IRS~16SW is a known binary with a period of 19.4 days, we only use literature $v_{z}$ measurements where the observation date is specified to the nearest day for this star \citep{Paumard2006, Zhu2008}. $v_{z}$ measurements from \citet{Paumard2001} are not used, as they derive their velocities from the He 2.058 $\mu$m line, which could potentially be biased by diffuse emission from the ISM \citep{Zhu2008}. Finally, we add a systematic error in quadrature of 3 km/s or 15 km/s to these $v_{z}$ values for WC or WN stars, respectively, to account for variations in the winds as described in Section~\ref{sect:wind_effects}. We include in our sample all stars with $v_z$ measurements from our work or the literature, for a total of 27 stars in our sample.

\section{Results}    
\label{sect:bin_search}
In this section, we first describe our method of identifying candidate spectroscopic WR binaries in Section~\ref{sect:bin_identification}. Section~\ref{sect:binary_cands} describes our new binary candidates and their possible physical properties, and section~\ref{sect:other_binary_cands} describes our observations of the previously known WR binaries in the Galactic center.

\subsection{Binary Candidate Identification}    
\label{sect:bin_identification}

\begin{figure}
\includegraphics[width=3.5in]{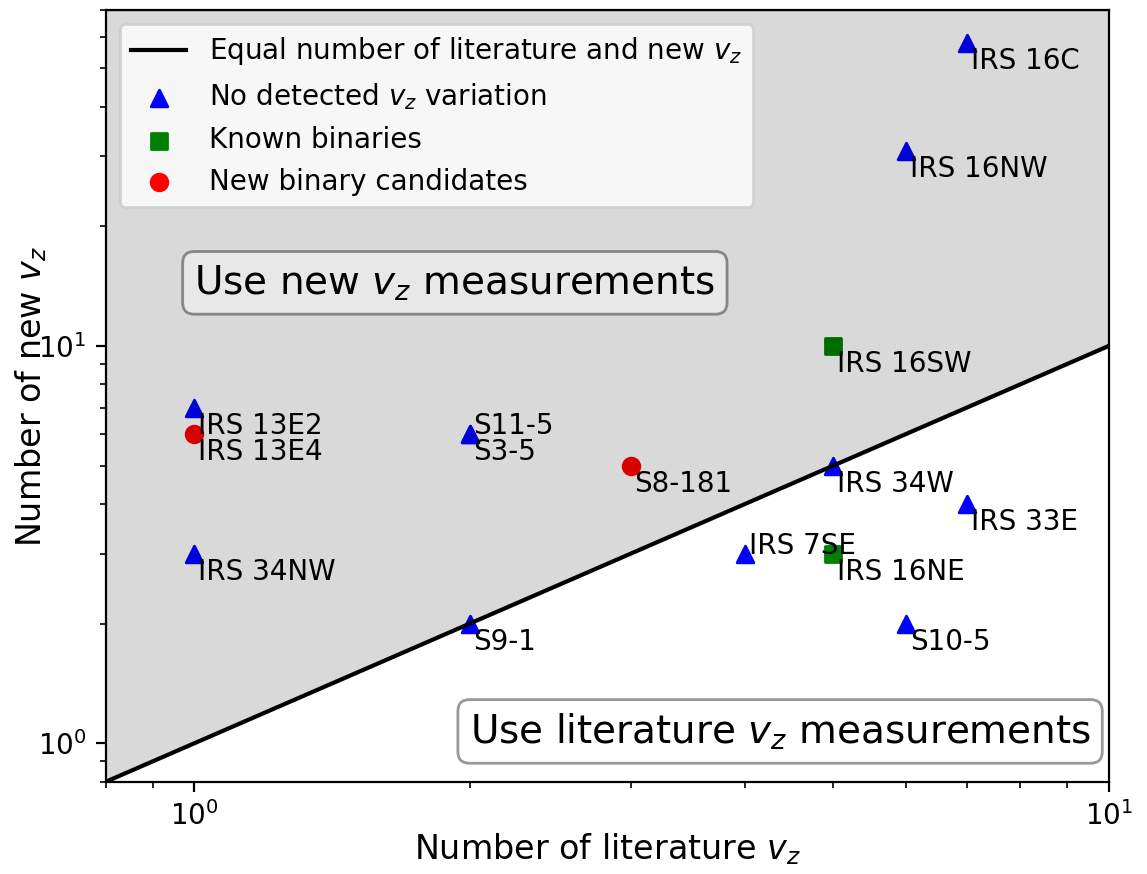}
\caption{\footnotesize Number of new relative $v_{z}$ measurements versus number of available literature $v_{z}$ measurements for stars which have at least two new $v_{z}$ values in our sample. We use new relative $v_{z}$ measurements in the binary search if there are a greater or equal number of them than measurements from the literature (the stars on the black line, or in the gray shaded region), and literature values otherwise (stars in the white region). We display new binary candidates as red circles, known binaries as green squares, and stars with no $v_{z}$ variation as blue triangles. The final number of $v_{z}$ values for each star used in the binary search is listed in Table \ref{tab:wr_table}. We do not mix literature and not $v_{z}$ measurements.}
\label{fig:new_vs_lit_rvs}
\end{figure}

We define binary candidates as stars which show at least 3$\sigma$ variation between any two pair of $v_{z}$ measurements: 
\begin{equation}
\frac{|v_{i} - v_{j}|}{\sqrt{\sigma_{i}^2 + \sigma_{j}^2}} > 3.0, 
\end{equation}
where $v_{i}$, $v_{j}$ and $\sigma_{i}$, $\sigma_{j}$ are the velocities and their uncertainties from epochs $i$ and $j$. These uncertainties include the wind-driven uncertainties from Section \ref{sect:wind_effects}. For stars with both literature and new $v_{z}$ values, we use our new relative $v_{z}$ measurements in the binary search when the number of our new $v_{z}$ measurements is greater than or equal to the number reported in the literature (14 stars,  Table~\ref{tab:wr_table}), and literature measurements otherwise (Figure \ref{fig:new_vs_lit_rvs}). We use literature  $v_{z}$ measurements for 9 stars where we do not have new measurements. We do not mix our new relative and literature absolute $v_{z}$ measurements for a particular star in the binary search, as we do not know the correction for the relative $v_{z}$ measurements to shift them to the absolute $v_{z}$ measurements from the literature. We remove changes in $v_{z}$ arising from the orbits of the stars around Sgr A* (\citet{Gillessen2017,vonFellenberg2022}, Bentley et al., in prep). For all but the closest stars (IRS~16C, IRS~16SW), the change in $v_{z}$ is expected to be less than 10 km/s, and will not change the results of this work.

\begin{figure*}
\begin{center}
\includegraphics[width=7in]{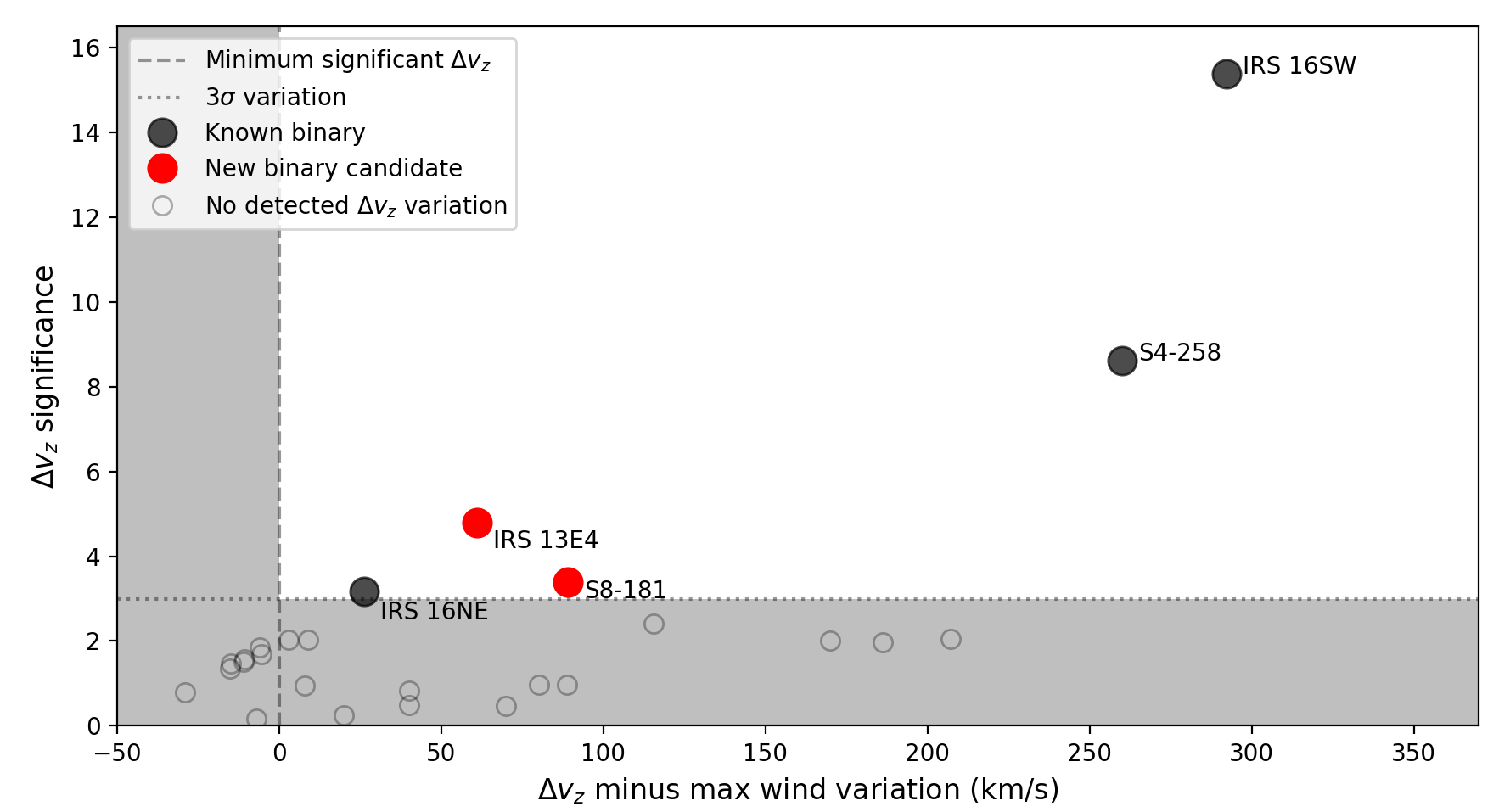}
\caption{\footnotesize $\Delta v_{z}$ significance versus $\Delta v_{z}$ minus the max wind variation for each spectral type \citep{Dsilva2020, Dsilva2022,Dsilva2023}, for the binary sample. Stars located in the white regions have $\sigma_{\Delta vz}$ that is not due to orbital motion about Sgr~A* and is greater than the scale of wind variability for WR stars. New binary candidates from this work are in red (IRS~13E4, S8-181), black filled stars are previously known binaries \citep{Ott1999, Tanner2006, Martins2006, Zhu2008, Pfuhl2014}, and black circles are stars without significant $v_{z}$ variation.}
\label{fig:wn_wc_variations}
\end{center}
\end{figure*}

We identify 5 binary candidates out of 27 WR stars in our sample: IRS~16SW, IRS~16NE, S4-258, IRS~13E4, and S8-181. IRS~13E4 and S8-181 are binary candidates reported here for the first time. Their relative $v_{z}$ measurements are displayed in Figure~\ref{fig:binary_rv_time}.  We plot the significance of the observed $\Delta v_{z}$ versus the magnitude of the $\Delta v_{z}$ for the WN $\&$ WC stars in Figure~\ref{fig:wn_wc_variations}. Binary candidates are listed in Table~\ref{tab:ls_table}. Below we discuss the details of each new candidate binary star, along with known binaries and candidates.

\subsection{Identified new candidate binaries} 
\label{sect:binary_cands}

\begin{figure*}[h]

        \centering
        \begin{tabular}{cc}
        \subfloat[S8-181 (WN8)]{\includegraphics[width=3in]{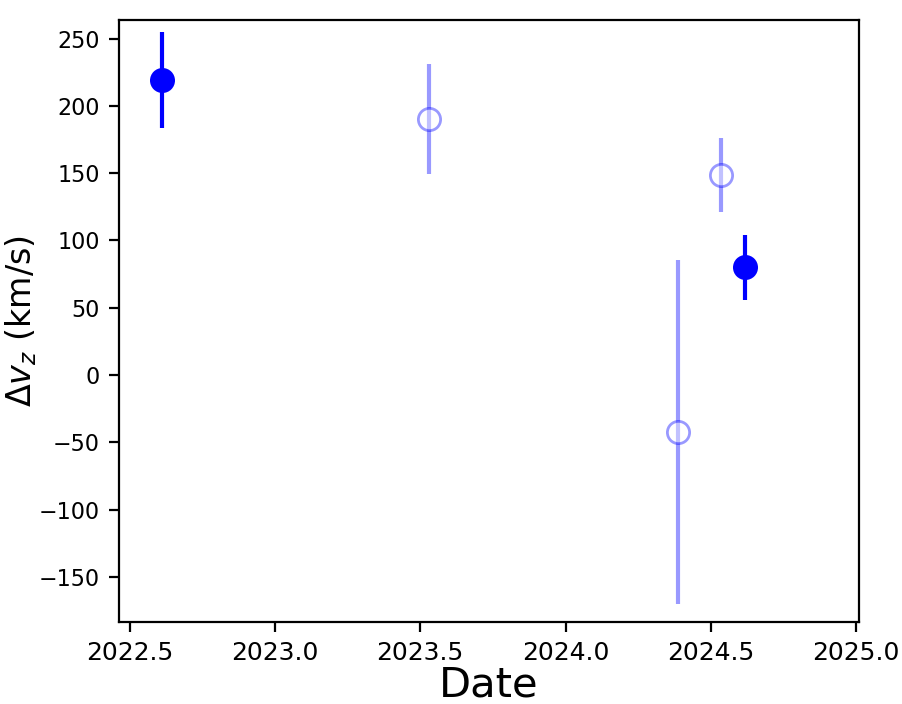}} &
        \subfloat[IRS~13E4 (WC9)]{\includegraphics[width=3in]{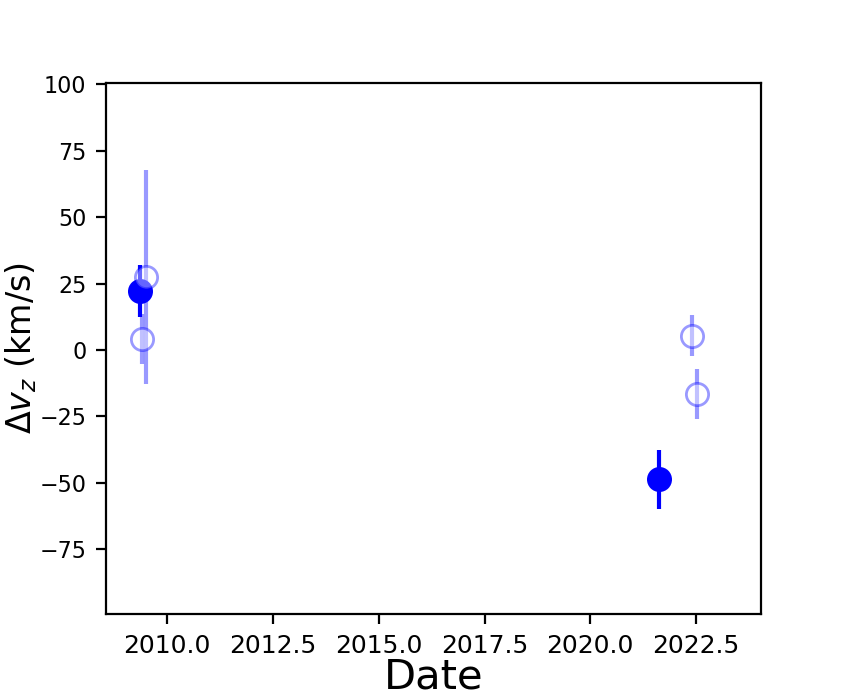}} \\
        \subfloat[IRS~13E2 (WN8)]{\includegraphics[width=3in]{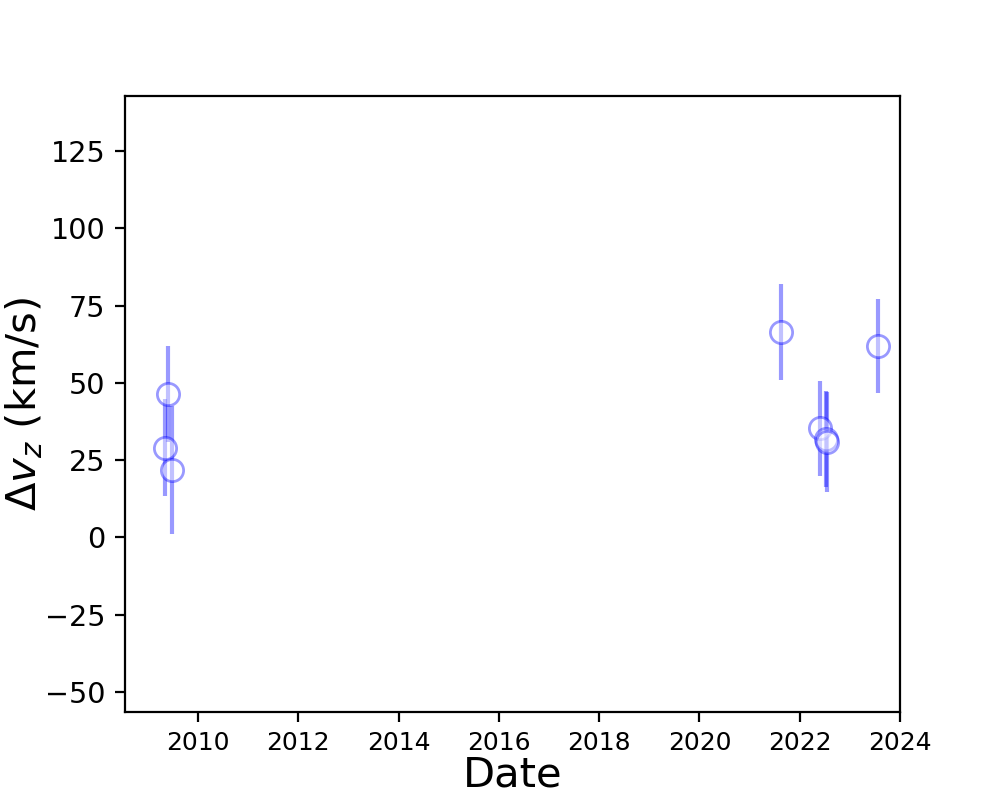}} &
        \subfloat[IRS~7SE (WC9)]{\includegraphics[width=3in]{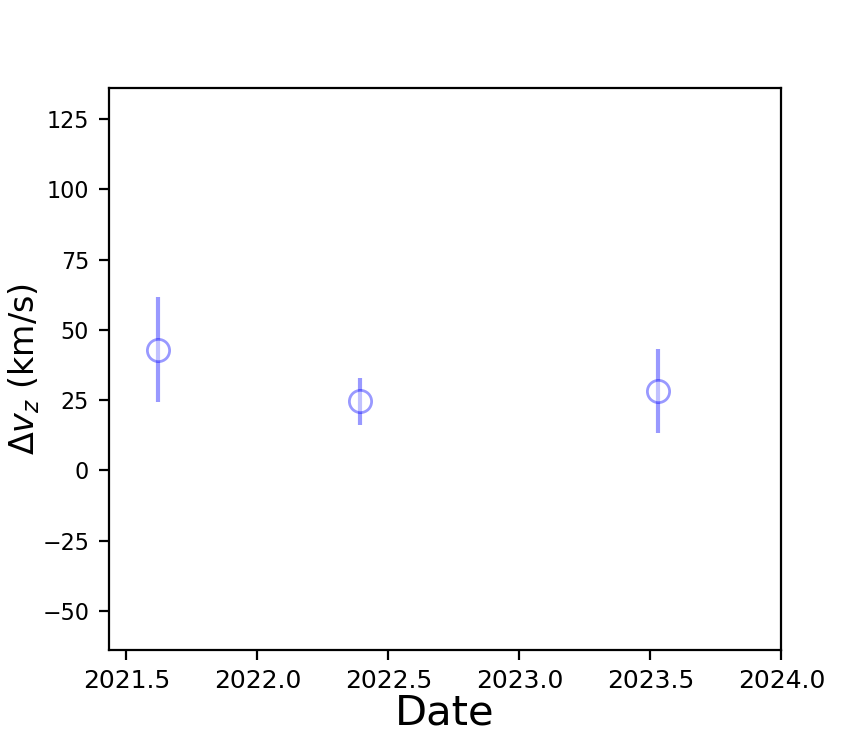}} \\
        \subfloat[IRS~16C (Ofpe/WN9)]{\includegraphics[width=3in]{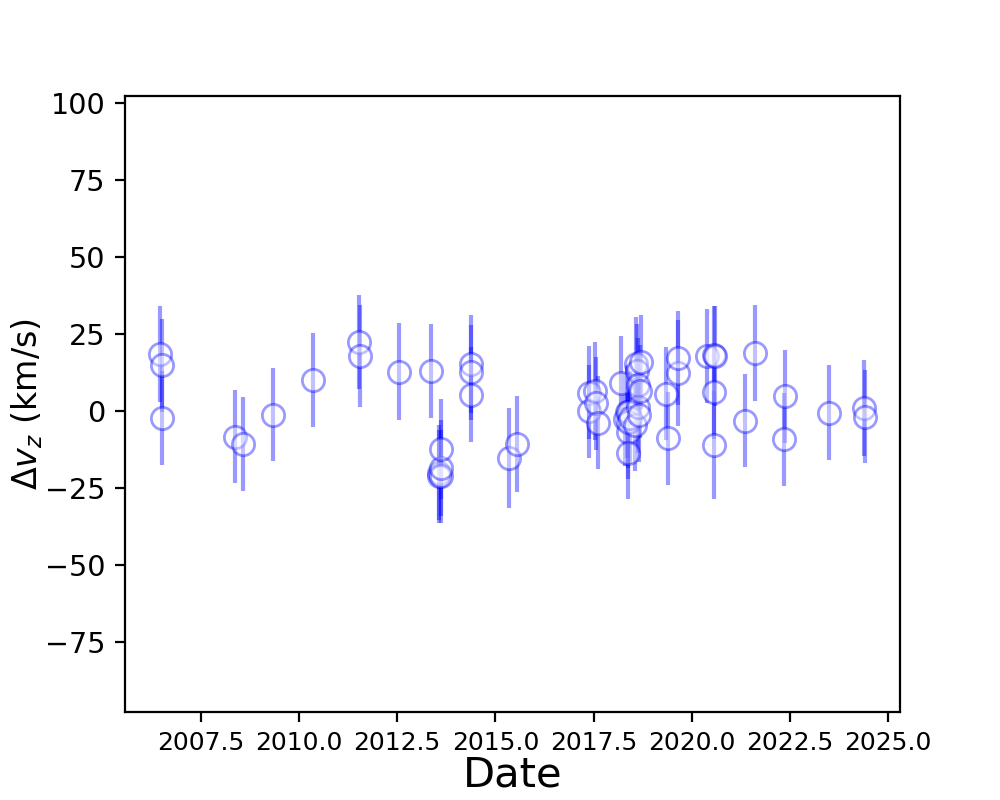}} &
        \subfloat[S3-5 (WN5/6)]{\includegraphics[width=3in]{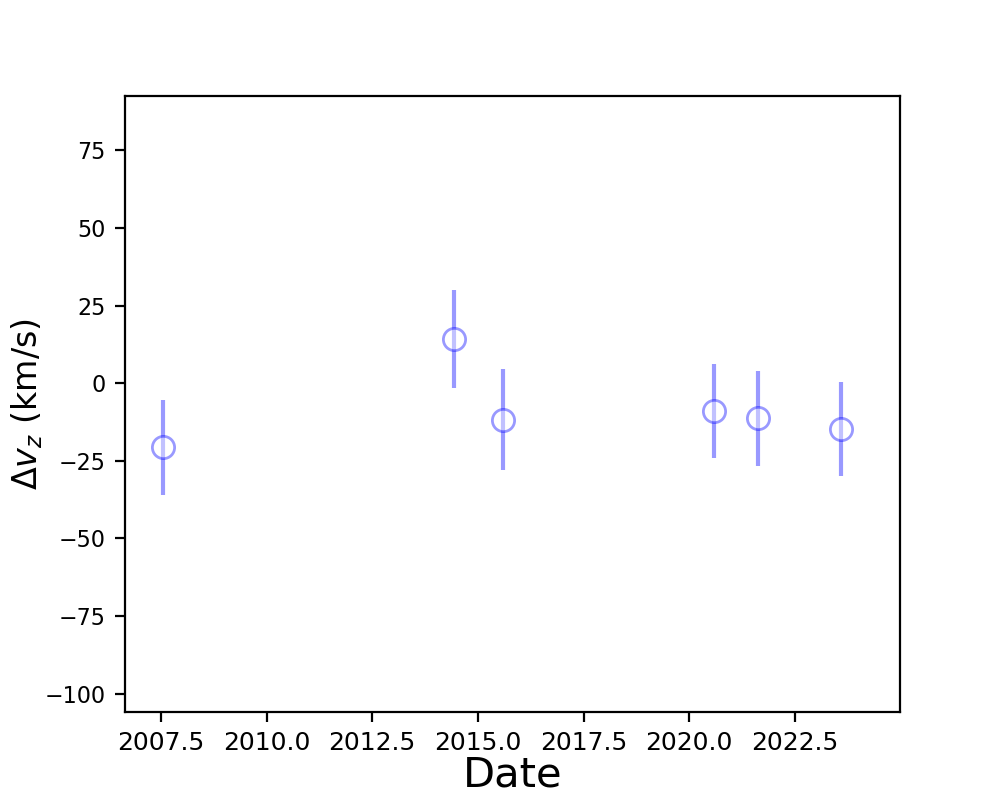}} \\
        \end{tabular}
        
\caption{$v_{z}$ shift from the template spectrum as a function of time for the binary candidates, along with two stars of the same spectral types as the candidates (WN8 for S8-181 and IRS~13E2, WC9 for IRS~13E4 and IRS~7SE), and two stars of other spectral types (Ofpe/WN9 for IRS~16C, WN5/6 for S3-5), which show no significant variation for reference. Filled points are the velocity shifts for the binary candidates with the most significant change between them, open points are the others. Acceleration due to IRS~16C's orbit around Sgr A* \citep{Gillessen2017, vonFellenberg2022} has been removed from our $v_{z}$ measurements.}
\label{fig:binary_rv_time}
\end{figure*}

\begin{figure*}
\includegraphics[width=7in]{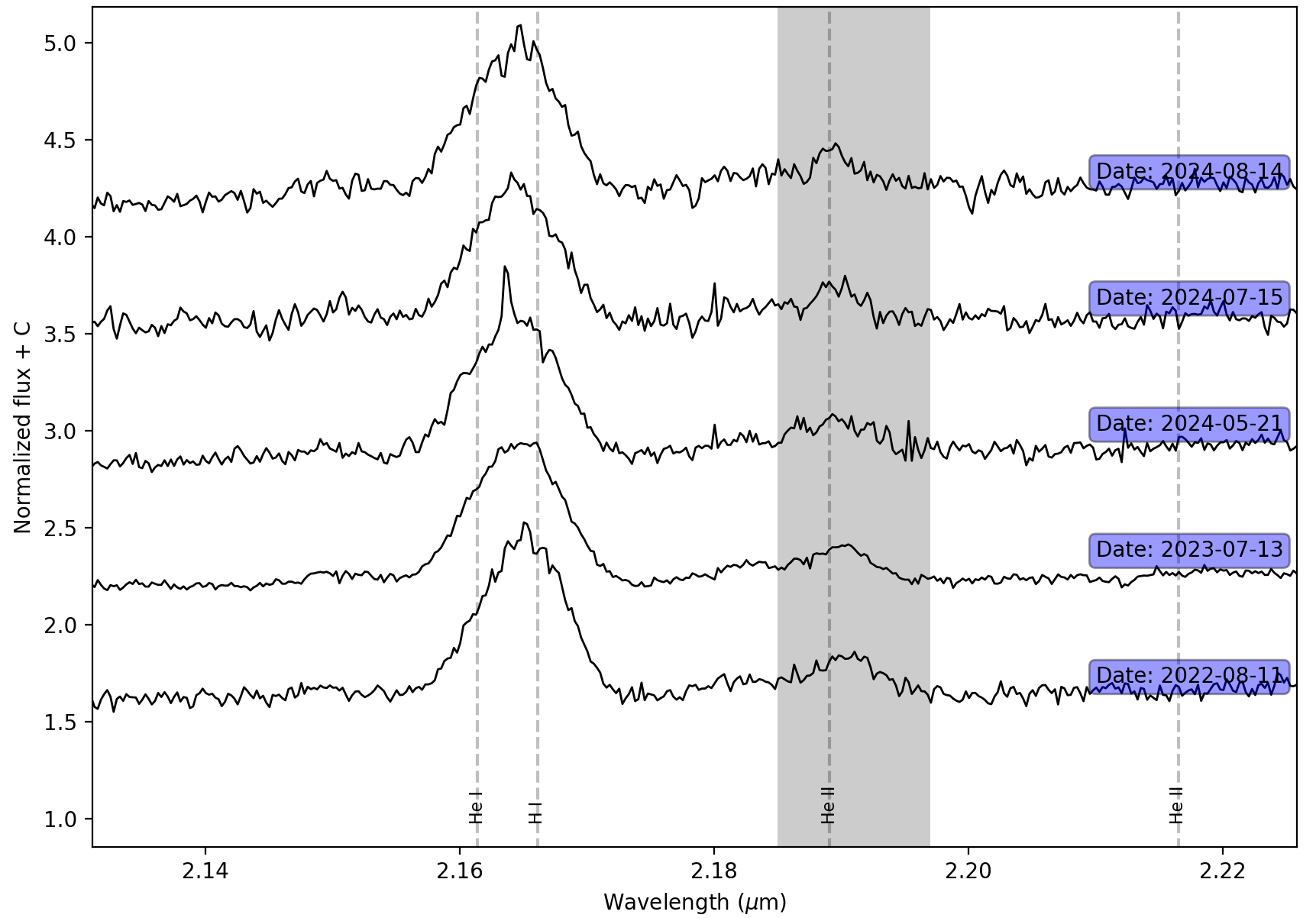}
\caption{\footnotesize Spectra of new binary candidate S8-181 used in this search, with spectral lines observed in WR star K-band spectra \citep{Figer1997} overlain. We use the He II lines at 2.189 $\mu$m to measure changes in $v_z$. He II lines in late-WN stars like S8-181 are noted to be lines the least effected by wind variability in \citet{Dsilva2023}. The region of the spectrum used in the cross-correlation is shaded in gray.}
\label{fig:s8-181_specs}
\end{figure*}

\paragraph{S8-181} 
\label{sect:s8-181_binary}
In our five spectra of this star (also known as AFNW, spectral type WN8, \citealt{Paumard2006, Feldmeier-Krause2015}, spectra shown in Figure~\ref{fig:s8-181_specs}), we see velocity variations of up to a 3.4$\sigma$ level (Figure~\ref{fig:binary_rv_time}, with the most significant variation being 139$\pm$42 km/s between our 2022-08-11 and 2024-08-14 spectra). Previous $v_{z}$ measurements of this star reported in \citet{Genzel2000}, \citet{Paumard2001, Paumard2006}, and \citet{Zhu2008} have not led to detections of any significant $v_{z}$ shifts, however, those $v_{z}$ measurements had a factor of 2 to 4 larger uncertainty than our measurements. Late-type WN stars such as S8-181 are known to show $v_{z}$ variation by up to 50 km/s because of these wind changes \citep{Dsilva2023}. Additionally, the changes in $v_{z}$ cannot be due to the gravitational influence of Sgr A*, as the maximum physically allowed change in $v_{z}$ from the black hole at the projected distance of S8-181 is $\sim$0.1 km/s over the time span of our $v_{z}$ measurements (two years, Tables \ref{tab:obs_table}, \ref{tab:rvs_table_all}). Because our observed $v_{z}$ variations are larger than these, we conclude that S8-181 is a binary candidate, but additional $v_{z}$ measurements are necessary to confirm its binary nature.

\begin{figure*}
\includegraphics[width=7in]{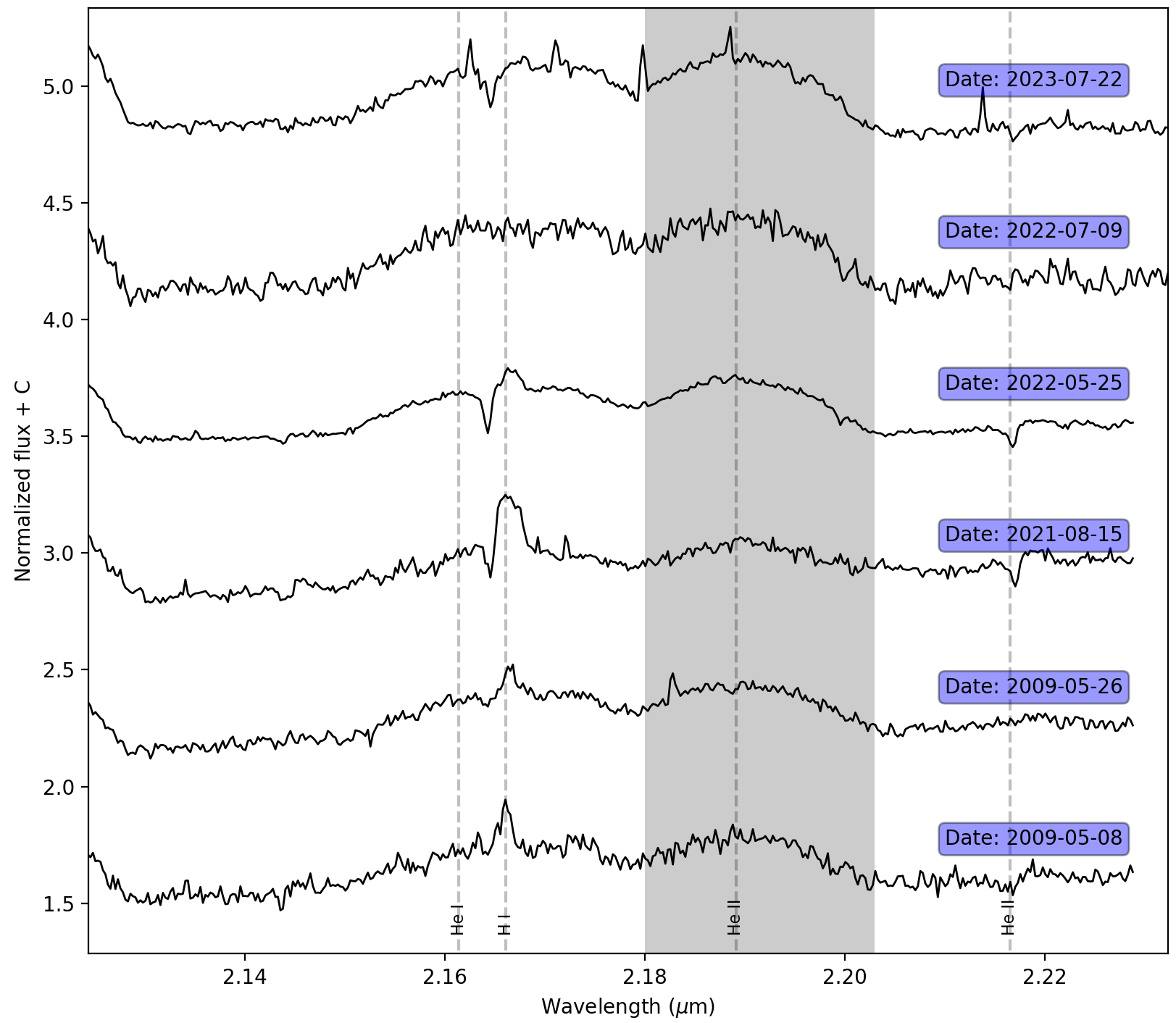}
\caption{\footnotesize Spectra of IRS~13E4 used in this search, with dashed vertical lines indicating spectral lines observed in WR star K-band spectra \citep{Figer1997} overlain. Our observed variability (71$\pm$15 km/s) is higher than expected for stars of this type, suggesting it may have a binary companion. \citet{Dsilva2020} found that He II lines in WC star spectra can show up to $\sim$20 km/s in wind variations. The region of the spectrum used in the cross-correlation is shaded in gray.}
\label{fig:irs13e4_specs}
\end{figure*}

\paragraph{IRS~13E4} 
\label{sect:irs13e4_binary}
For this star (spectral type WC9, \citet{Paumard2006}, spectra shown in Figure~\ref{fig:irs13e4_specs}), we detect a $v_{z}$ variation of 4.8$\sigma$ significance (Figure~\ref{fig:binary_rv_time}, with the most significant variation being 71$\pm$15 km/s between our 2009-05-08 and 2021-08-15 spectra). \citet{Dsilva2020} found WC stars like IRS~13E4 (spectral type WC9, \citet{Paumard2006}) show less wind-driven $v_{z}$ shifts than WN stars, with variability reaching up to 10 km/s, much less than our observed shift. 

We see several features in the spectra disappear between 2022-05-25 and 2022-07-09 observations (and seen in Figure~\ref{fig:irs13e4_specs}), which are contaminants from the surrounding IRS 13 complex, and are visible in the background spectra plotted in Appendix \ref{sect:irs13e4_background}. These features are a P-Cygni-like feature superimposed on top of the He I/He II lines near 2.163 $\mu$m \citep{Figer1997} which is also visible in the IRS~13E4 spectrum in \citet{Martins2007}, and another P-Cygni-like emission and absorption feature near 2.218 $\mu$m, possibly associated with the 2.2165 $\mu$m He II line and also seen in the \citet{Martins2007} spectrum. As these features are not present in observations taken in the smaller 20 mas plate scale on OSIRIS. Increasing the radius of the annulus used to determine the background spectra in the 20 mas observations to match the radii used with the 35 mas and 50 mas plate scales caused the features to reappear, indicating they are not from the star itself. Changes in the aperture size does not significantly affect the $v_{z}$ measurement. Increasing the aperture radius used in each observation by 50$\%$ and redetermining the $v_{z}$ measurements did not change any of the values by more than 1$\sigma$. The observed $v_{z}$ variation cannot again be attributed to the influence of Sgr A*. At the projected distance of IRS 13E4, the maximum physically allowed change in $v_{z}$ over the time span of our $v_{z}$ measurements (14 years, Tables \ref{tab:obs_table}, \ref{tab:rvs_table_all}) is $\sim$5 km/s. Again, because our $v_{z}$ variations are larger than both the allowed acceleration from Sgr A* and the wind variations, we are attributing it to binary motion.

\subsection{Previously known binaries and binary candidates} 
\label{sect:other_binary_cands}
\paragraph{IRS~16SW}

\begin{figure}
\includegraphics[width=3.5in]{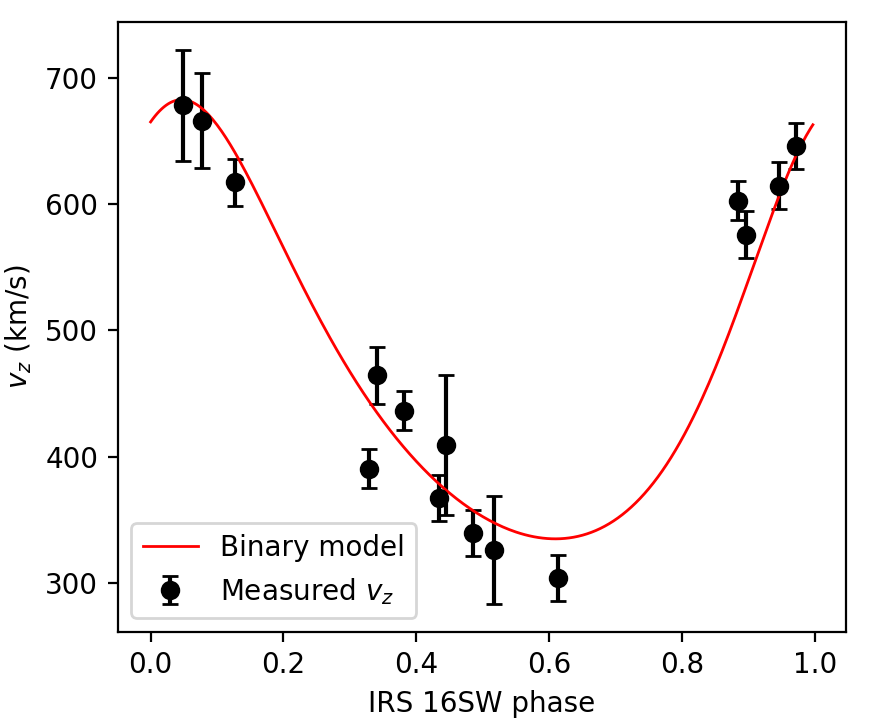}
\caption{\footnotesize Measured IRS~16SW $v_{z}$ values compared to the IRS~16SW binary model from \citet{Martins2006} and \citet{Peeples2007}, folded onto the period of the binary. We can see that the previously determined model for the binary (parameters from \citet{Martins2006}, with P, $T_{0}$ from \citet{Peeples2007}) agrees very well with the determined $v_{z}$ values from this work.}
\label{fig:irs16sw_folded_rvs}
\end{figure}

The first binary identified in the Galactic center was IRS~16SW (primary and secondary are both spectral type Ofpe/WN9, \citet{Ott1999,Depoy2004,Rafelski2007, Paumard2006, Martins2006, Peeples2007, Feldmeier-Krause2015}). We find very large $v_{z}$ variation at a $\sim$15$\sigma$ level, and when we account for the $a_{z}$ acceleration and fold the $v_{z}$ measurements onto the known period of the binary (19.4 days, \citet{Rafelski2007, Martins2006, Peeples2007}), our $v_{z}$ measurements are in good agreement with the expected values. IRS~16SW has a 0.31 AU semi-major axis, with a likely 50 $M_{\odot}$ primary and an equal-mass companion \citep{Martins2006, Peeples2007}. We compare the $v_{z}$ measurements of IRS~16SW folded onto the period of the binary from \citet{Peeples2007} in Figure~\ref{fig:irs16sw_folded_rvs}, and show the similarities between the $v_{z}$ measurements from this work and the binary parameters determined in \citet{Martins2006} and \citet{Peeples2007}.

\paragraph{IRS~16NE}
\label{sect:irs16ne_binary}
IRS~16NE (primary spectral type Ofpe/WN9, \citet{Paumard2006, Feldmeier-Krause2015}) was identified as a binary candidate in \citet{Tanner2006}, and confirmed in \citet{Pfuhl2014}. We detect significant $v_{z}$ variation as with IRS~16SW, at a $\sim$ 3.2$\sigma$ level. Based on \citet{Pfuhl2014}, IRS~16NE has a 224 day period, 1.2 AU semi-major axis, likely 50 $M_{\odot}$ primary, and a companion with a likely mass ratio between 0.6 and 1. We were only able to obtain 3 relative $v_{z}$ measurements for this star, and our observations were all taken at phases of the binary orbit where the primary star does not experience significant $v_{z}$ variation. Thus, we use the literature $v_{z}$ measurements from \citet{Zhu2008} and \citet{Paumard2006} when calculating the most significant $v_{z}$ variation.

\paragraph{S4-258} 
\label{sect:S4-258_binary}

S4-258 (spectral type WN7?, \citealt{Paumard2006, Feldmeier-Krause2015}) was first identified as an eclipsing binary in \citet{Pfuhl2014}. We use data points from \citet{Pfuhl2014} for this star, which only lists the phase of the binary system (folded on the 2.276 day period from photometry), and not the actual dates of observation. 

\begin{deluxetable*}{lllll}[ht]
\tablecolumns{5} 
\tablewidth{0pc}
\footnotesize
\tablecaption{Binary Search Results\label{tab:ls_table}}
\tablehead{ 
	\colhead{Name} &
    \colhead{Most significant $\Delta v_{z}$} &
    \colhead{$\Delta v_{z}$ significance} &
	\colhead{$N_{v_{z}}$} &
	\colhead{Binary?} \\
    \colhead{} &
    \colhead{km/s} &
    \colhead{} &
	\colhead{} &
	\colhead{}  
}

\startdata
Binaries+candidates & & & & \\
IRS 16SW & 343$\pm$22 & 15.4 & 15 & Known binary \\ 
S4-258 & 310$\pm$36 & 8.6 & 12 & Known binary \\
IRS 13E4 & 71$\pm$15 & 4.8 & 6 & New binary \\ 
S8-181 & 139$\pm$42 & 3.4 & 5 & New binary \\ 
IRS 16NE & 76$\pm$24 & 3.2 & 5 & Known binary \\
\hline
Non-detections & & & & \\
IRS 9W & 165$\pm$70 & 2.4 & 5 & -- \\
IRS 34NW & 53$\pm$26 & 2.1 & 2 & -- \\ 
S9-9 & 220$\pm$110 & 2.0 & 5 & -- \\
IRS 33E & 59$\pm$29 & 2.0 & 7 & -- \\ 
S6-95 & 200$\pm$100 & 2.0 & 6 & -- \\
S9-114 & 104$\pm$53 & 2.0 & 3 & -- \\ 
IRS 16C & 44$\pm$24 & 1.8 & 58 & -- \\ 
S9-283 & 220$\pm$120 & 1.8 & 2 & -- \\ 
IRS 13E2 & 45$\pm$26 & 1.7 & 8 & -- \\ 
IRS 16NW & 39$\pm$25 & 1.6 & 31 & -- \\ 
S11-5 & 39$\pm$26 & 1.5 & 5 & -- \\ 
IRS 34W & 35$\pm$24 & 1.4 & 5 & -- \\ 
S3-5 & 35$\pm$26 & 1.3 & 6 & -- \\ 
S6-93$^{a}$ & 140$\pm$140 & 1.0 & 5 & -- \\
IRS 16SW-E & 90$\pm$90 & 1.0 & 2 & -- \\
IRS 7SE & 18$\pm$19 & 0.9 & 3  & -- \\
S6-90 & 90$\pm$110 & 0.8 & 2 & -- \\
S10-5 & 21$\pm$27 & 0.8 & 5 & -- \\ 
S10-136 & 50$\pm$100 & 0.5 & 2 & -- \\
IRS 29N & 80$\pm$180 & 0.4 & 2 & -- \\
S2-16 & 30$\pm$120 & 0.3 & 2 & -- \\
S9-1 & 3$\pm$18 & 0.2 & 2 &  -- \\
\enddata

\tablenotetext{a}{Discussed as a potential binary in \citet{Paumard2001} and \citet{Pfuhl2014}, but only shows 1$\sigma$ significant variations, and as such does not meet our criteria for being a binary candidate.}

\end{deluxetable*}

\section{Discussion}    
\label{sec:discussion}
In this section, we use our new binary star candidates to infer the intrinsic binary fraction ($f_{bin}$) of the WR stars in the Galactic center (Section~\ref{sect:binary_frac}), and compare our $f_{bin}$ to those derived in previous studies of the binary population of the Galactic center and WR star binaries in the Milky Way field (Section~\ref{sect:fbin_comp}).

\subsection{Binary fraction of WR stars in the Galactic center} 
\label{sect:binary_frac}
If we assume that all binary candidates identified in this work are truly binaries, then there are 5 binaries within our detection threshold out of the 27 WR star sample. We use a simulation to assess the sensitivity of our experiment and infer a binary fraction ($f_{bin}$) of 0.56$\pm$0.18, given the number of binaries detected and assuming the orbits are random and isotropically distributed. Each simulation involves 4 steps: (1) determining which stars are binaries, (2) assigning physical properties of the generated binaries, (3) generating $v_{z}$ values and uncertainties based on the binary orbits and the observations of the stars, and (4) finally, testing the generated $v_{z}$ measurements with the same statistical test of $v_{z}$ variation significance for binary candidates. Each step is described in more detail below.

\begin{enumerate}
\item We start by taking the 27 WR stars with multiple $v_{z}$ measurements from this work and the literature, and randomly assigning whether a star is a binary or not based on a specified $f_{bin}$. The 1000 simulations each were repeated for 101 $f_{bin}$ values between 0 and 1, and we then determined the median and standard deviation in $f_{bin}$ for different numbers of detected $v_{z}$ variables.

\item If the star is a binary, we then generate a binary companion for that star. For all binary companions, $\omega$, $T_{0}$, and cos($i$) are each given a random value selected from a uniform distribution. The eccentricity and mass ratio are randomly drawn from the \citet{Sana2012} distributions. If the star is a WN, WN/WC, or Ofpe/WN9 type star, then the orbital period of the companion is generated using the \citet{Sana2012} distribution, which \citet{Dsilva2022, Dsilva2023} found to agree with the period distributions of early and late-type WN stars. For WC stars, we use the period distribution of WC binaries from \citet{Dsilva2022}, which favors longer period systems. The mass-ratio assumptions are the same as in the binary candidate period/semi-major axis limit calculations in Section~\ref{sect:bin_identification}. We note that the distribution of binary orbital properties could be different in the Galactic center compared to the young clusters and field stars used in \citet{Sana2012, Dsilva2022, Dsilva2023}, but these observed distributions nonetheless provide reasonable guesses as to what they could be at the Galactic center. We discuss our assumptions on WR binary physical properties further in Appendix~\ref{sect:fbin_assumptions}. 

\item The $v_{z}$ measurements of each binary system are calculated at the times of their actual observations (or the first day of the month, if only the month is given), with the number of $v_{z}$ measurements in the simulations equaling the number for each star used in this work (Table \ref{tab:wr_table}). The uncertainty on a specific $v_{z}$ measurement in the simulation is equal to the real uncertainty on the measurement on that associated date (e.g. 16 km/s uncertainty is used for the 2006-06-18 observation of IRS 16C in the simulation, Table \ref{tab:rvs_table_all}). 

\item We then determine the highest significance of any $v_{z}$ variation in the same manner used to detect binary candidates, and count the number of detected binaries with a significance greater than 3$\sigma$. 

\end{enumerate}

\begin{figure}
\includegraphics[width=3.5in]{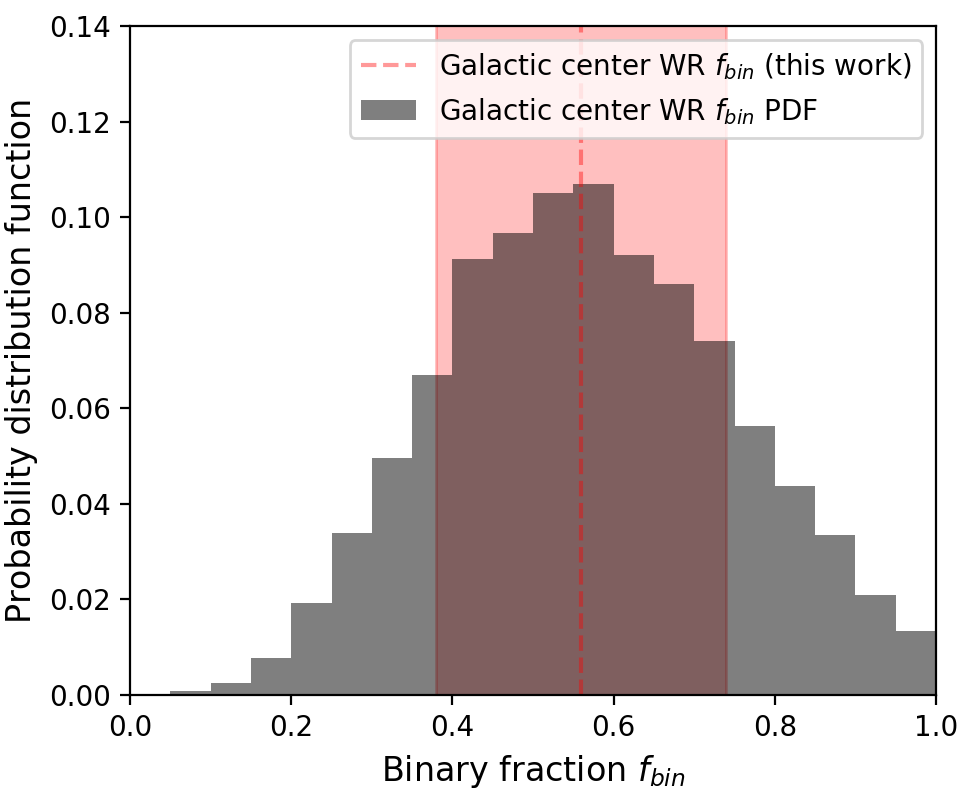}
\caption{\footnotesize Posterior distribution function (PDF) for $f_{bin}$, along with our  $f_{bin}=0.56\pm0.18$ for detecting 5 binaries, is shown in grey. Our lower limit points to a high binary fraction for the Galactic center WR stars, comparable to the high binary fractions for field O and WR stars \citep{Sana2012,Dsilva2022,Dsilva2023,Offner2023} and agreeing with the lower limit from photometry for the Galactic center OB and WR population \citep{Gautam2024}.}
\label{fig:fbin_pdf}
\end{figure}

 Based on these simulations, we infer that a spectroscopic detection of 5 binaries (IRS~16SW, IRS~16NE, S4-258, S8-181, IRS~13E4) out of 27 stars corresponds to $f_{bin}=0.56\pm0.18$ (Figure~\ref{fig:fbin_pdf}). When the posterior distribution functions of $f_{bin}$ from this work and \citet{Gautam2024} are combined, we find $f_{bin}=0.69\pm0.17$ for the broader young stellar population.

As WN and WC stars may have different binary properties \citep{Dsilva2022,Dsilva2023}, and may be compared to their respective field populations separately, we split the populations into two groups based on spectral type. WN/WC hybrid stars are included in the WN sample to be consistent with our binary search. We also do not split the WN population into early and late types, as the two appear to have similar distributions in binary properties \citep{Dsilva2022,Dsilva2023}. 

We note that our conclusions about the binary fractions depend on several factors. This includes the assumptions in the sensitivity simulations about the binary physical properties such as mass ratio and period distributions as we cannot independently constrain these parameters (and they have not been well constrained for WR stars in general, \citet{Dsilva2020,Dsilva2022,Dsilva2023,Deshmukh2024,Shenar2024}). A significant factor in our uncertainty in the binary fraction is also the number of measurements and their precision. Our inference about binaries is based on $v_{z}$ variations, but larger numbers of measurements will be needed to confirm their binary nature and to measure important parameters like their periods. Furthermore, the radial velocity measurements can be improved with better models of their atmospheres and line profiles. 

Additionally, some of the late-type WC stars in the Galactic center which produce dust (for example, IRS 29N) have been suggested to be colliding wind binaries \cite{Rafelski2007,Feldmeier-Krause2015}. The sources of dust from many dust-producing WR stars is an ongoing topic of research, but many of the dust-producing WR stars appear to be colliding wind binaries \citep[e.g.][]{Crowther2007,Tuthill2008,Sander2012,Williams2014,White2026}. One of our new binary candidates, IRS 13E4, is a dust-producing WC9 star \citep{Martins2007}, but we do not detect velocity variations for the other dust producing WR stars. Since we do not detect velocity variations for the other dust-producing WR stars, and there are dust-producing WR stars which appear to be single \citep{Sander2012,Dsilva2020}, we are conservatively not using dust-production alone as an indicator of binarity and do not consider them binary candidates in the study.

\subsection{Comparison of the Galactic center WR binary fraction to other works}    
\label{sect:fbin_comp}

\begin{figure*}
\includegraphics[width=7in]{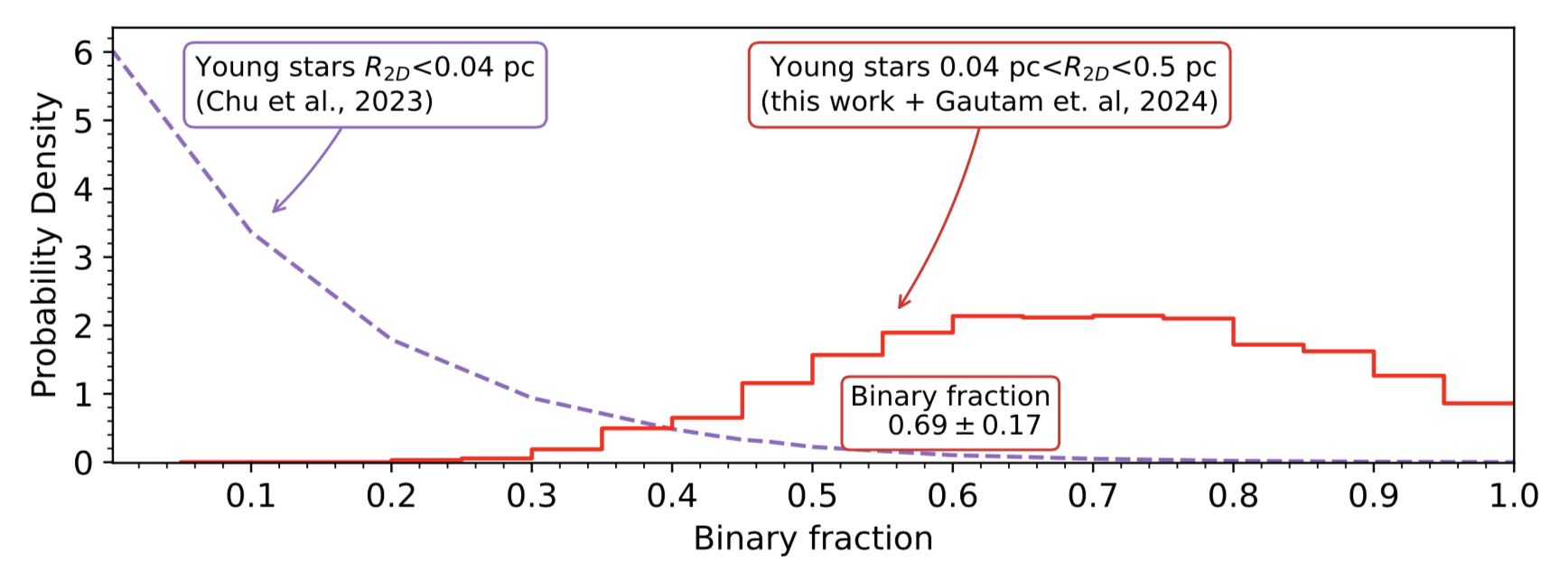}
\caption{\footnotesize PDF for our combined $f_{bin}$ from this work and \citet{Gautam2024}, compared to the PDF from \citet{Chu2023} of the binary fraction of the young stars within the central 0.04~pc in projection (finding $f_{bin}$ $\le$0.17 at 68$\%$ confidence). We find our $f_{bin}$ is very different from the \citet{Chu2023} values, supporting the idea of a radial dependence of $f_{bin}$ in the Galactic center. Figure adapted from \citet{Gautam2024}. All PDFs are normalized so their integral is equal to 1.}
\label{fig:fbin_pdf_compare}
\end{figure*}

The binary fraction in this work is consistent with previous observations at the Galactic center of massive young stars \citep[e.g.][]{Pfuhl2014, Tanner2005,Peeples2007,Zhu2008,Rafelski2007,Chu2023,Gautam2019,Gautam2024}.  \cite{Gautam2024} found that $f_{bin}$ $\ge$0.72 at the 68$\%$ level from photometric observations of the young stellar population. In addition, \citet{Pfuhl2014} reported a spectroscopic binary fraction of $0.30^{+0.34}_{-0.21}$ at 95$\%$ confidence in their study targeting bright OB and WR stars in the Galactic center. Though they did not convert this to an intrinsic binary fraction, their observed spectroscopic binary fraction is close within uncertainties to the observed spectroscopic binary fractions found in other young clusters, which have an intrinsic binary fraction $\sim70\%$ \citep[e.g.][]{Sana2012,Sana2013, Ritchie2022, Clark2023}. Our measurement of a binary fraction $f_{bin}=0.56\pm0.18$ (Table~\ref{tab:fbin_combo_table}), is consistent with both these studies. \citet{Naze2012} conducted a search for X-ray counterparts the spectral type Ofpe/WN9 stars (included in our WN sample as discussed in Section~\ref{sect:wr_sample}) in the Galactic center, and did not detect any X-ray sources coincident with the Ofpe/WN9 stars which resembled a colliding wind binary, suggesting that we are not missing any close binaries with strong colliding winds which have not previously been detected. When compared with the studies from \citet{Gautam2024} and \citet{Chu2023} (which detected no binaries at $r<$ 0.04~pc from Sgr A*), we have further evidence that the binary fraction is higher beyond 0.04~pc from the black hole as proposed by \citet{Gautam2024} (Figure~\ref{fig:fbin_pdf_compare}). We note that there are recent tentative suggestions by \citet{Peissker2024} that there are binaries within 0.04 pc, but these objects appear to be more unusual stars (possibly merger products of much lower mass stars, \citealt{Ciurlo2020}) than our sample. 

Our measured $f_{bin}$ is also consistent with recent studies of the binary fraction for other evolved massive stars in the Milky Way \citep[e.g,][]{Sana2012, Dsilva2020, Dsilva2022, Mahy2022,Dsilva2023, Shenar2024, Deshmukh2024,Offner2023}. \citet{Dsilva2020,Dsilva2022,Dsilva2023} determined the binary fractions for four subsets of the WR population in the Milky Way field (WC, early WN, late WN, and early+late WN) through a similar spectroscopic search to our own. For WN stars, the Galactic center $f_{bin}=0.56\pm0.18$ (Table~\ref{tab:fbin_combo_table}) agrees with the field value for WN stars $f_{bin,WN}=0.52^{+0.15}_{-0.12}$ from \citet{Dsilva2023}. \citet{Dsilva2023} do caution that their study could miss longer period WN binaries, which would not be missed in the WC sample due to their smaller wind-driven $v_{z}$ variations, and thus their period distribution would be preferentially weighted towards shorter period systems and their $f_{bin}$ underestimated. If that is the case, our $f_{bin}$ would also be an underestimate as we use the same period distribution. However, any biases would likely affect both samples equally due to our similar binary identification methodology, and so the true $f_{bin}$ values would still agree. Our $f_{bin}$ is lower than \citet{Dsilva2022,Shenar2024}'s WC binary fraction lower limit of $f_{bin,WC}>0.74$. The smaller number of WC stars in our sample combined with many of our $v_{z}$ measurements for WC stars having $\sim100$ km/s uncertainties (Table \ref{tab:rvs_table_all}), means our constraints on the Galactic center $f_{bin}$ mainly come from WN stars, and so the disagreement is not surprising. Additionally, the difference in WN and WC binary fractions from \citet{Dsilva2020,Dsilva2022,Dsilva2023} was disputed by \citet{Deshmukh2024}, who attributed it to a small sample size and noted examples of short period ($\le$100 days) WC binaries. Our period and semi-major axis limits, along with the scale of observed $v_{z}$ variation, are within the range of properties and observed $v_{z}$ variation for known WC binaries in both samples (\citet{Dsilva2020, Deshmukh2024} and references within), so our result agrees with both works. We also note that the Galactic center WR star binary fraction is very similar field O-star and LBV populations \citep{Sana2012, Mahy2022}, which are thought to be WR star progenitors. Our combined binary fraction of $f_{bin}=0.69\pm0.17$ from this work and \citet{Gautam2024} is also in excellent agreement with the binary fraction of O-stars in young Milky Way clusters from \citet{Sana2012} of $f_{bin}=0.69\pm0.09$.

\begin{deluxetable*}{cccc}
\tablecolumns{4} 
\tablewidth{0pc} 
\tablecaption{Comparison of Galactic center and field young star binary fractions\label{tab:fbin_combo_table}}
\tablehead{ 
	\colhead{Population} &
	\colhead{Location} & 
	\colhead{Radial range (pc)} & 
	\colhead{$f_{bin}$} 
}
\startdata
This work (Galactic center WN+WC stars) & Further from Sgr A* & 0.065-0.45 & 0.56$\pm$0.18 \\ 
Galactic center WR and OB stars $\ge$1" of Sgr A* (\citet{Gautam2024}, 68$\%$ conf.) & Further from Sgr A* & 0.04-0.3 & $\ge$0.72 \\ 
Galactic center WR and OB stars $\ge$1" of Sgr A* (this work + \citet{Gautam2024}, 68$\%$ conf.) & Further from Sgr A* & 0.04-0.45 & 0.69$\pm$0.17 \\ 
Galactic center B-stars $\le$1" of Sgr A* (\citet{Chu2023}, 68$\%$ conf.) & Closest to Sgr A* & $\le$0.04 & $\le0.17$ \\
\hline
O-stars in nearby young clusters (\citet{Sana2012}, 68$\%$ conf.) & Milky Way field & -- & 0.69$\pm$0.09 \\
Nearby WC stars (\citet{Dsilva2022,Shenar2024}, 68$\%$ conf.) & Milky Way field & -- & $>0.74$ \\
Nearby WNL+WNE stars (\citet{Dsilva2023,Dsilva2022}, 68$\%$ conf.) & Milky Way field & -- & $0.52^{+0.15}_{-0.12}$ \\
\enddata
\end{deluxetable*}

\section{Conclusion}    
\label{sect:conclusion}

We present the longest time-baseline study yet of the WR stars withing 0.5~pc of the Galactic center, reporting 146 new spectroscopic observations from 2006-2024 combined with literature spectroscopic observations from 1994-2024. We present 146 new radial velocities, which are used in a search for new binary candidates to constrain the binary fraction of the Galactic center WR stars ($f_{bin}$). 

Our binary search targeted 27 WR stars throughout the Galactic center young stellar cluster, making it the largest sample size yet (more than three times the number of WR stars compared to the previous spectroscopic study in \citet{Pfuhl2014}) in a spectroscopic study of the cluster, and identified 2 new binary candidates (S8-181, IRS~13E4). Based on the observed changes in radial velocity and mass estimates based on their spectral types, we identified potential orbital period ranges for the binaries, and find they are comparable to typical periods for WR binaries in the field. We find a binary fraction $f_{bin}=0.56\pm0.18$, in agreement with the $f_{bin}$ lower limit from photometric monitoring of the larger young stellar population from \citet{Gautam2024}, and when combined with their result, we find a binary fraction of $f_{bin}=0.69\pm0.17$. Our result is also consistent with the suggestion from \citet{Gautam2024} that $f_{bin}$ increases beyond the central 0.04~pc, where the binary fraction of the young stars appears to be $>0.17\%$ at 68$\%$ confidence \citep{Chu2023,Gautam2024}.

\section{Acknowledgments}
\label{sect:acknowl}

We thank the staff of the Keck Observatory and the Gemini Observatory, especially Sherry Yeh, Carlos Alvarez, Randy Campbell, Jim Lyke, Tony Connors, John Pelletier, Max Piper, Julie Renaud-Kim, and Arina Rostopchina for all their help in obtaining the new observations.  We thank the anonymous referee for their comments and suggestions. We also thank Smadar Naoz for her comments and input. A.M.G. acknowledges support from her Lauren B. Leichtman and Arthur E. Levine Endowed Astronomy Chair. Support for this work was provided by the Gordon and Betty Moore Foundation under award No. 11458, and by National Science Foundation award No. 1909554. Additional support was received from the UCLA Galactic Center Star Society and the Brinson Prize Fellowship (held by M. W.H.). The international Gemini Observatory, a program of NSF NOIRLab, is managed by the Association of Universities for Research in Astronomy (AURA) under a cooperative agreement with the U.S. National Science Foundation on behalf of the Gemini Observatory partnership: the U.S. National Science Foundation (United States), National Research Council (Canada), Agencia Nacional de Investigaci\'{o}n y Desarrollo (Chile), Ministerio de Ciencia, Tecnolog\'{i}a e Innovaci\'{o}n (Argentina), Minist\'{e}rio da Ci\^{e}ncia, Tecnologia, Inova\c{c}\~{o}es e Comunica\c{c}\~{o}es (Brazil), and Korea Astronomy and Space Science Institute (Republic of Korea). The W. M. Keck Observatory is operated as a scientific partnership among the California Institute of Technology, the University of California, and the National Aeronautics and Space Administration. The authors wish to recognize that the summit of Maunakea has always held a very significant cultural role for the indigenous Hawaiian community. We are most fortunate to have the opportunity to observe from this mountain. The Observatory was made possible by the generous financial support of the W. M. Keck Foundation.

\bibliography{atc_bib}

\section{Appendix} 
\label{sect:appendices}

\subsection{IRS~13E4 background}    
\label{sect:irs13e4_background}
\begin{figure*}
\includegraphics[width=7in]{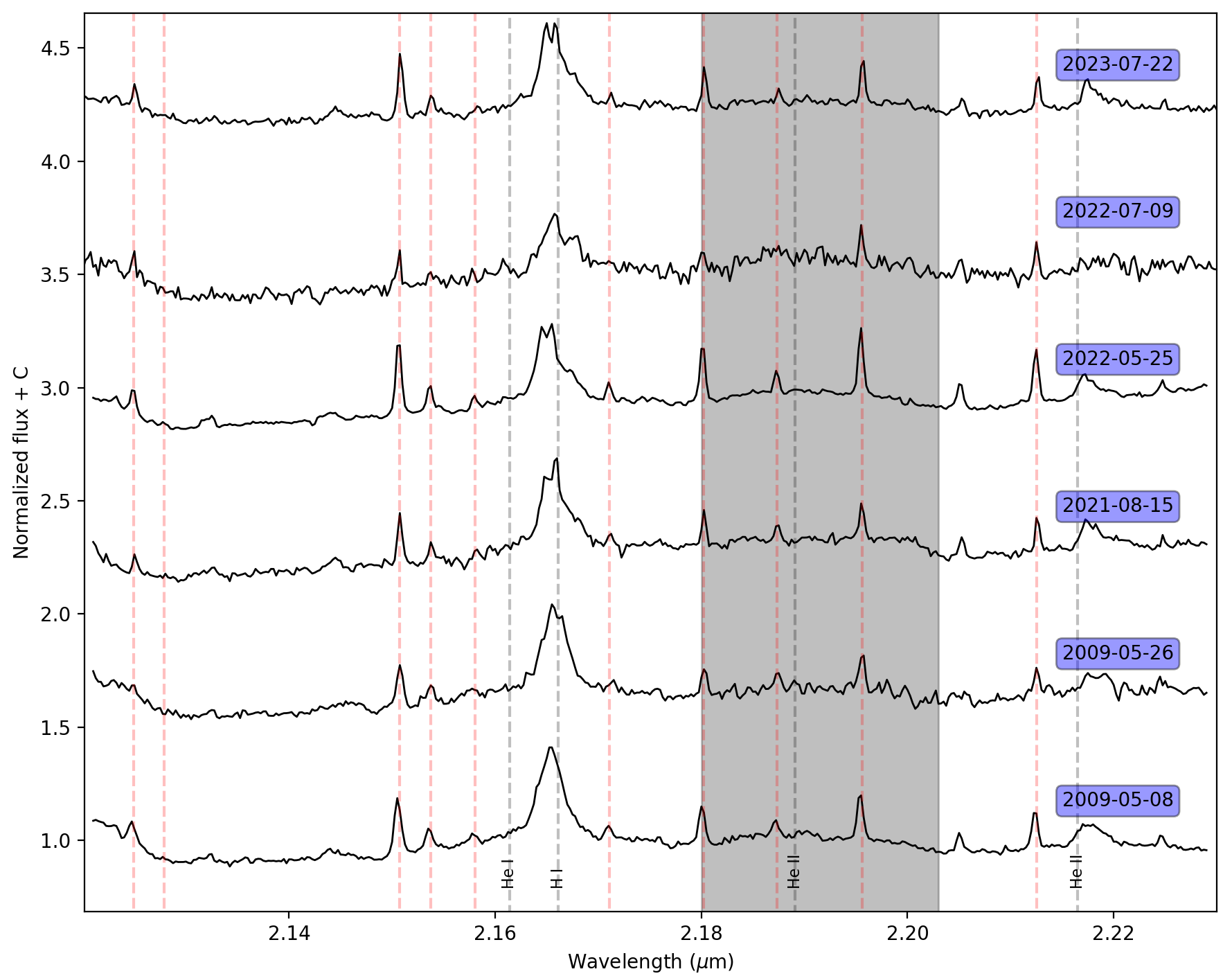}
\caption{\footnotesize The background spectra used to remove sky and background features in our IRS~13E4 spectra. The gray dashed lines indicate spectral lines observed in WR star K-band spectra \citep{Figer1997}, the red dashed lines indicate OH sky lines in our spectra, and the gray shaded region is the wavelength range used for cross-correlation for IRS~13E4 (2.18,2.203 $\mu$m, Table \ref{tab:rvs_table_all}). Although there is contamination from nearby IRS~13E2 and surrounding gas near Brackett-$\gamma$ and the 2.217 $\mu$m He II line, these are outside the regions of the spectra used to obtain $v_{z}$ measurements for this star, and thus will not effect our results.}
\label{fig:irs13e4_bgs_skylines}
\end{figure*}

One of our binary candidates, IRS~13E4, is a member of the IRS~13E cluster of massive stars, including another WR star in our sample, IRS~13E2. Its location within this region of high stellar density makes it especially susceptible to contamination by other nearby sources, such as IRS~13E2. In Figure \ref{fig:irs13e4_bgs_skylines}, we plot the background spectra extracted around IRS~13E4 used to remove sky and background features in each spectrum. The method of background construction and extraction is described in \citet{Do2009,Do2013}. The spectra in Figure \ref{fig:irs13e4_bgs_skylines} do not show any contamination from nearby sources in the wavelength range used for cross-correlation for IRS~13E4 (2.18,2.203 $\mu$m, Table \ref{tab:rvs_table_all}).

\subsection{Assumptions on the correction from observed to intrinsic $f_{bin}$}    
\label{sect:fbin_assumptions}
To correct our observed $f_{bin}$ into an intrinsic $f_{bin}$, we have to make assumptions on the underlying distributions of binary properties, in particular the mass-ratio, eccentricity, and different period distributions for the WN and WC populations. In order to assess how our assumptions on these parameters affect our $f_{bin}$, we tried varying the distributions (all of which are expressed as power-laws in the manner of \citet{Sana2012}) to see how they change our result.

For the mass-ratio distribution (expressed by probability density pdf(q) $\sim$ $q^{\kappa}$), \citet{Dsilva2022} found that changes in $\kappa$ did not substantially change their results. We assumed a nearly flat density ($\kappa$ = -0.1$\pm$0.6) from \citet{Sana2012} to match the distribution of the the WR stars' likely O-star progenitors. Repeating our $f_{bin}$ calculations with $\kappa$ = -0.7 yielded $f_{bin}$=0.75$\pm$0.17, while using a massive secondary-heavy $\kappa$ = 0.5 yielded $f_{bin}$=0.51$\pm$0.16. These only shift our $f_{bin}$ measurement by 1$\sigma$ at most. Unfortunately, $\kappa$ for WR stars is poorly constrained, and so future measurements may lead to revisions of our $f_{bin}$ measurements. However, some binary evolution models such as \citet{Langer2020} favor WR stars being more massive than their companions, with $\kappa \le 0$, which would favor a slightly higher $f_{bin}$ than what we find with $\kappa$ = -0.1.

For the eccentricity distribution (with probability density pdf(e) $\sim$ $e^{\eta}$), we again assume the \citet{Sana2012} O-star value ($\eta$ = -0.45) to match the progenitors, although note that interactions between the stars could lead to the older WR binaries, in particular close binaries, having lower eccentricity orbits. We calculated $f_{bin}$ with $\eta$ = 1.0 and $\eta$ = -1.0, and found lower limits of $f_{bin} \ge 0.73$ and $f_{bin} \ge 0.67$ respectively for the two $\eta$ values. As these are both close to what we found for our assumed $\eta$, we say that our conclusions do not significantly depend on our assumptions on $\eta$.

We use the \citet{Sana2012} period distribution (probability density pdf(P) $\sim$ $P^{\pi}$) for WN stars as it is consistent with the \cite{Dsilva2022, Dsilva2023} $\pi$ values for both early and late type WN stars, but for WC stars, we use $\pi$ = 1.90 based on \citet{Dsilva2022}. \citet{Deshmukh2024} argued that the very long period-heavy WC distribution from \citet{Dsilva2022} could be due to small sample size missing short period binaries. To account for this, we again determined $f_{bin}$ now using the short-period heavy \citet{Sana2012} $\pi$ = -0.55 for all WR stars, and found $f_{bin}=0.50\pm0.16$, consistent with our value and with the \citet{Dsilva2022,Dsilva2023} distributions. Considering WC stars alone, a we find $f_{bin,WC}=0.50\pm0.25$.

\subsection{Spectroscopic observations and $v_{z}$ measurements}    
\label{sect:obs_and_rvs}

Table \ref{tab:obs_table} lists all the new spectroscopic observations in this work, and Table \ref{tab:rvs_table_all} lists all the $v_{z}$ measurements used in this work.
 
\startlongtable
\begin{deluxetable*}{ccccccclc}
\tablecolumns{9} 
\tablewidth{0pc}
\tabletypesize{\scriptsize}
\tablecaption{Summary of Spectroscopic Observations\label{tab:obs_table}}
\tablehead{ 
	\colhead{Date} &
	\colhead{$N_{frames}$} &
	\colhead{WR spectra} &
	\colhead{Instrument} &
	\colhead{Filter} &
	\colhead{I. Time/Frame} &
	\colhead{Plate Scale} &
	\colhead{Stars observed} &
	\colhead{Sgr A* Offset (RA,Dec)$^{1}$} \\
    \colhead{(UT)} &
	\colhead{} &
    \colhead{} &
    \colhead{} &
	\colhead{} &
    \colhead{s} &
    \colhead{as/pix} &
    \colhead{} &
	\colhead{"} 
}
\startdata
2006-06-18 & 9 & 2 & OSIRIS & Kn3 & 900 & 0.035 & {\tiny IRS 16C, IRS 16SW} & 0.0,0.0 \\ 
2006-06-30 & 9 & 2 & OSIRIS & Kn3 & 900 & 0.035 & {\tiny IRS 16C, IRS 16SW} & 0.0,0.0 \\ 
2006-07-01 & 9 & 2 & OSIRIS & Kn3 & 900 & 0.035 & {\tiny IRS 16C, IRS 16SW} & 0.0,0.0 \\ 
2007-07-18 & 10 & 2 & OSIRIS & Kn3 & 900 & 0.035 & {\tiny IRS 16SW-E, S3-5} & 2.30,-0.22 \\ 
2007-07-20 & 11 & 1 & OSIRIS & Kn3 & 900 & 0.035 & {\tiny IRS 34W} & -1.96,0.88 \\ 
2008-05-16 & 11 & 3 & OSIRIS & Kn3 & 900 & 0.035 & {\tiny IRS 16C, IRS 16SW, IRS 16NW} & 0.0,0.0 \\
2008-06-03 & 11 & 1 & OSIRIS & Kn3 & 900 & 0.035 & {\tiny S3-12} & 1.32,-1.88 \\ 
2008-06-10 & 5 & 1 & OSIRIS & Kn3 & 900 & 0.035 & {\tiny IRS 16NE} & 2.89,0.98 \\ 
2008-07-25 & 11 & 2 & OSIRIS & Kn3 & 900 & 0.035 & {\tiny IRS 16C, IRS 16SW} & 0.0,0.0 \\ 
2009-05-05 & 12 & 1 & OSIRIS & Kn3 & 900 & 0.035 & {\tiny IRS 16SW} & 0.0,0.0 \\ 
.. & .. & .. & .. & .. & .. & .. & .. & .. \\ 
\enddata
\tablenotetext{1}{Offset of the center of the instrument field of view. For entries where the pointing was not listed, the pointing was at the location in Table \ref{tab:wr_table} of the stars observed.}
\tablenotetext{}{(This table is available in its entirety in machine-readable form.)}
\end{deluxetable*}
\startlongtable
\begin{deluxetable*}{lllrcccccccr}
\tablecolumns{12} 
\setlength{\tabcolsep}{1pt}
\tabletypesize{\scriptsize}

\tablecaption{Radial Velocities used in this work\label{tab:rvs_table_all}}
\tablehead{ 
	\colhead{Name} &
	\colhead{$\Delta v_{z}$} &
	\colhead{Date} &
	\colhead{MJD} &
	\colhead{Line used} &
	\colhead{fit $\lambda$ range} &
	\colhead{S/N} &
	\colhead{Instrument$^{a}$} & 
	\colhead{Resolution} &
	\colhead{Filter} &
	\colhead{Plate Scale} &
	\colhead{Source} \\
    \colhead{} &
	\colhead{km/s} &
    \colhead{} &
    \colhead{} &
    \colhead{} &
    \colhead{$\mu$m} &
    \colhead{} &
	\colhead{} &
    \colhead{} &
    \colhead{} &
	\colhead{"/pix} &
	\colhead{} 
}
\startdata
New $\Delta v_{z}$s & & & & & & & & & & & \\
IRS 16C & 87$\pm$17 & 2006-06-18 & 53904.41 & Br$\gamma$ & 2.16,2.175 & 53 & OSIRIS & 4000 & Kn3 & 0.035 & This work \\ 
IRS 16C & 66$\pm$17 & 2006-06-30 & 53916.38 & Br$\gamma$ & 2.16,2.175 & 67 & OSIRIS & 4000 & Kn3 & 0.035 & This work \\ 
IRS 16C & 83$\pm$17 & 2006-07-01 & 53917.35 & Br$\gamma$ & 2.16,2.175 & 112 & OSIRIS & 4000 & Kn3 & 0.035 & This work \\ 
IRS 16C & 54$\pm$17 & 2008-05-16 & 54602.5 & Br$\gamma$ & 2.16,2.175 & 110 & OSIRIS & 4000 & Kn3 & 0.035 & This work \\ 
IRS 16C & 51$\pm$17 & 2008-07-25 & 54672.33 & Br$\gamma$ & 2.16,2.175 & 67 & OSIRIS & 4000 & Kn3 & 0.035 & This work \\ 
IRS 16C & 58$\pm$17 & 2009-05-06 & 54957.54 & Br$\gamma$ & 2.16,2.175 & 160 & OSIRIS & 4000 & Kn3 & 0.035 & This work \\ 
IRS 16C & 67$\pm$17 & 2010-05-08 & 55324.52 & Br$\gamma$ & 2.16,2.175 & 80 & OSIRIS & 4000 & Kn3 & 0.035 & This work \\ 
IRS 16C & 74$\pm$17 & 2011-07-10 & 55752.36 & Br$\gamma$ & 2.16,2.175 & 64 & OSIRIS & 4000 & Kn3 & 0.035 & This work \\ 
IRS 16C & 70$\pm$18 & 2011-07-19 & 55761.33 & Br$\gamma$ & 2.16,2.175 & 27 & OSIRIS & 4000 & Kn3 & 0.035 & This work \\ 
IRS 16C & 62$\pm$17 & 2012-07-22 & 56130.3 & Br$\gamma$ & 2.16,2.175 & 42 & OSIRIS & 4000 & Kn3 & 0.035 & This work \\ 
.. & .. & .. & .. & .. & .. & .. & .. & .. & .. & .. & .. \\ 
\enddata
\tablenotetext{a}{Instrument locations are as follows: W. M. Keck Observatory (OSIRIS, NIRSPEC), Gemini-North (NIFS), ESO VLT (SPIFFI/SINFONI, ISAAC), Canada-France-Hawaii Telescope (BEAR), ESO-MPG (MPE-3D).}
\tablenotetext{b}{\citet{Pfuhl2014} only reported $v_{z}$ measurements and orbital phases for their S4-258 observations, along with the plate scale and resolution ranges for their SINFONI observations. Specific observation dates and the associated plate scales and resolutions were not reported.}
\tablenotetext{}{(This table is available in its entirety in machine-readable form.)}
\end{deluxetable*}

\end{document}